\documentclass[lettersize,journal]{IEEEtran}
\usepackage{amsmath,amsfonts}
\usepackage{algorithmic}
\usepackage{algorithm}
\usepackage{array}
\usepackage{textcomp}
\usepackage{stfloats}
\usepackage{url}
\usepackage{verbatim}
\usepackage{graphicx}
\usepackage{cite}
\usepackage{amssymb}  
\usepackage{multirow}
\usepackage{arydshln}
\usepackage{graphicx}
\usepackage{subfigure}
\usepackage{ntheorem}
\usepackage{color}
\newtheorem{definition}{\textbf{Definition}}
\newtheorem{assumption}{\textbf{Assumption}}
\newtheorem{lemma}{\textbf{Lemma}}

\newtheorem{theorem}{\textbf{Theorem}}
\newtheorem{remark}{\textbf{Remark}}

\newtheorem{corollary}{\textbf{Corollary}}

\begin{document}

\title{Zero-determinant Strategy for Moving Target Defense: Existence, Performance, and Computation}


\author{Zhaoyang Cheng, Guanpu Chen, \IEEEmembership{Member,~IEEE}, Yiguang Hong, \IEEEmembership{Fellow,~IEEE}, Ming Cao, \IEEEmembership{Fellow,~IEEE}, \\ and Mikael Skoglund, \IEEEmembership{Fellow,~IEEE}
\thanks{Z. Cheng, G. Chen, and M. Skoglund are with the School of Electrical Engineering and Computer Science, KTH Royal Institute of Technology, Stockholm, Sweden.
        {\tt\small zhcheng,guanpu,skoglund@kth.se}}
\thanks{Y. Hong is with  the  Department  of  Control  Science  and  Engineering  and Shanghai  Research  Institute  for  Intelligent  Autonomous Systems, Tongji University, Shanghai, China.
{\tt\small yghong@iss.ac.cn}}%
 \thanks{M. Cao is with the Institute of Engineering and Technology, University of Groningen, Groningen, The Netherlands.
		{\tt\small m.cao@rug.nl}}
}



\maketitle

\begin{abstract}

Moving Target Defense (MTD) is commonly formulated as a repeated security game to mitigate persistent threats. Although the strong Stackelberg equilibrium (SSE) characterizes the defender’s optimal strategy in the leader-follower framework, computing the SSE often incurs high computational complexity, which significantly limits its practical deployment in MTD problems with multiple targets. This paper proposes adopting a zero-determinant (ZD) strategy for constructing an MTD strategy that achieves both high defensive performance and substantially low computational complexity. We first derive a necessary and sufficient condition for the existence of ZD strategies and investigate the performance of ZD strategies, which shows their upper-bound performance matches that of the SSE strategy. We then formulate two programs to find the optimal ZD strategy parameters under different conditions.
Moreover, we design an algorithm to compute the proposed ZD strategies, along with the computational complexity analysis in comparison with the traditional SSE computation. Finally, we conduct experiments on two practical applications to verify our results.

\end{abstract}


\section{Introduction}

 Moving target defense (MTD) has emerged as a powerful approach for enhancing system security to deal with increasingly persistent threats \cite{zhuang2014towards,feng2017signaling,zhou2021sdn}. Unlike static protection mechanisms, MTD enables the defender to proactively migrate protection among multiple critical targets or update the system configuration
over time, thereby preventing attacks 
\cite{hong2015assessing,sengupta2020survey}.
These interactions are naturally captured by a repeated security game formulation, where players adopt their strategies in response to historical outcomes. Due to its dynamic and proactive nature, MTD has been widely studied and deployed in cyber–physical systems (CPS), including the Internet of Things (IoT) networks, cloud computing environments, and operating systems \cite{yue2025non,shen2024sgd3qn,celdran2024rl}.
Accordingly, how to effectively establish MTD strategies becomes an increasingly core problem.

The interactions between the defender and the attacker are commonly modeled by a leader–follower framework, since defensive strategies are strategically committed and infrequently updated after deployment \cite{yang2025herd,xu2024consistency,fang2017coordinated}. The attacker is usually a follower choosing the best response (BR) strategy after exploring and adapting to defensive strategies, while the defender, as a leader, aims to maximize its utility considering
this attacker. The equilibrium is defined as the strong Stackelberg equilibrium (SSE) \cite{korzhyk2011stackelberg,vorobeychik2012computing,9722864}, and the SSE defines the defender’s optimal strategy in this framework.  Thus, the defender's SSE strategy is often considered for MTD deployment to achieve high defense performance \cite{feng2017stackelberg,li2019optimal,li2022robust,qian2022ei}.

However, the practical deployment of the SSE strategy in repeated security games is severely hindered by its high computational complexity. Even in one-shot Stackelberg security games, computing the SSE, which is reformulated as a mixed-integer linear program, grows rapidly in complexity with the number of targets, due to the integer constraints \cite{korzhyk2011stackelberg}. When extended to repeated security games, the problem becomes even more challenging to formulate as a mixed-integer non-convex program, substantially complicating the search for a globally optimal and exact solution \cite{vorobeychik2012computing,basu2023complexityBB}.  Therefore, although the SSE strategy is optimal for the defender, the computational complexity imposes a heavy burden, preventing the defender from adopting effective MTD strategies in a timely manner \cite{basu2023complexity,basu2025information}.  This motivates our exploration to find a new approach for the defender's MTD strategy with both the sufficiently high profit and significantly low complexity.

Zero-determinant (ZD) strategies have attracted increasing attention due to their remarkable capability of unilateral utility enforcement, originally introduced in the iterated prisoner’s dilemma (IPD) \cite{press2012iterated}. A ZD strategy allows a player to unilaterally enforce a linear relationship between the expected utilities of the interacting players, regardless of the opponent’s strategy. This unilateral enforcement enables ZD-driven methods to operate in an open-loop manner, thereby facilitating rapid deployment, which makes them attractive in human–computer interaction (HCI) and evolutionary games \cite{wang2016extortion,hilbe2013evolution}.

Recently, ZD strategies have also drawn growing attention in CPS, including IoT systems, crowdsourcing systems, and blockchain platforms. However, most existing studies focus on two-action games \cite{hu2019solving,chen2022strategic,tang2024cooperation,feng2025zero} or restrict attention to equalizer ZD strategies \cite{wang2019moving}, both of which represent only a limited subset of the ZD strategy space. In security games, two-action games correspond to scenarios with only two feasible targets, while equalizer ZD strategies fix only one player's utility, thereby limiting their ability to fully explore diverse strategic interactions in general MTD. These limitations highlight the need for a general and computationally efficient ZD-driven approach tailored to MTD with multiple targets.

In this paper, we are motivated to develop a ZD-driven approach for constructing MTD strategies that achieve high defensive performance and maintain low computational complexity, in contrast to traditional SSE strategies. To this end, we aim to investigate the existence of ZD strategies and characterize their performance. Further, an efficient algorithm should be designed to compute the proposed ZD strategies, and the complexity needs to be analyzed.

The main contributions are summarized as follows:

\begin{itemize}
    \item We formulate ZD strategies for repeated security games as MTD strategies and characterize their key properties. In particular, we derive a necessary and sufficient condition for the existence of ZD strategies (Theorem~\ref{th::1}). We further investigate the performance of ZD strategies and show that their upper-bound performance matches that of the SSE strategy  (Theorem~\ref{le::ZDlemma2}).

\item We develop programs to find key linear parameters of ZD strategies through simplified constraints. If the upper-bound performance is achievable, we propose a program to find a set of ideal ZD strategy parameters that align with the defender’s SSE utility (Theorem~\ref{th::2}). If not, we formulate the other program to find the optimal ZD strategy parameters that maximize the defender’s utility (Theorem~\ref{th::optimal_ZD}).

\item  We design an algorithm (Algorithm~\ref{Al::ZD construction}) to compute the proposed ZD strategies and formally show its optimality guarantee (Theorem~\ref{pro::algorithmZD}). Further, we show that the proposed algorithm achieves substantially lower computational complexity than traditional SSE computation.

\end{itemize}

\textbf{Related work:} In the following, we provide literature reviews from the perspectives of SSE and ZD strategies, focusing on their application in CPS, especially MTD problems.

\textit{SSE strategy:} The SSE has been widely applied in CPS-related security scenarios. In one-shot Stackelberg security games, SSE is commonly used to characterize the defender’s optimal protection strategy \cite{korzhyk2011stackelberg,sinha2018stackelberg}. A single-leader multi-follower (SLMF) Stackelberg framework was proposed to enhance physical-layer security via coordinated multiple relays \cite{fang2017coordinated}. Signaling games have also been employed to model deception and information leakage in security settings \cite{pawlick2018modeling}. To account for both misperception and deception, a hypergame framework was developed for SLMF security games \cite{9722864}. More recently, tri-level Stackelberg models have been introduced to analyze insider influence in hierarchical cybersecurity systems \cite{xu2024consistency}. Overall, these studies demonstrate the effectiveness of SSE in modeling strategic interactions in CPS.

SSE has attracted significant attention in MTD problems, particularly in settings with repeated interactions between defenders and attackers. To capture the dynamic nature of MTD mechanisms, SSE-based approaches have been extended using Markov decision models \cite{feng2017stackelberg}. Subsequent studies further investigated temporal scheduling and spatio-temporal deployment of MTD strategies \cite{li2019optimal,li2020spatial}. In addition, practical factors such as signaling \cite{feng2017signaling} and deception \cite{nguyen2019deception} have been incorporated to influence attacker behavior in repeated security games. More recently, robust Stackelberg formulations were proposed to address uncertainty in attacker responses within two-player Markov games \cite{li2022robust}. Stackelberg-based MTD approaches have also been applied to practical systems, such as edge intelligence \cite{qian2022ei}. Overall, SSE plays a central role in the design and analysis of MTD strategies and has been extensively adopted in existing studies.

\textit{ZD strategy:} ZD strategies were first introduced in the IPD \cite{press2012iterated}, showing that a player adopting a ZD strategy can unilaterally enforce a linear relationship between the expected utilities of both players, which were extensively studied in public goods games, HCI, and evolutionary games \cite{wang2016extortion,govaert2020zero,hilbe2013evolution}. Focusing on the enforced linear relationship, ZD strategies were broadly classified into three typical types, including equalizer strategies, which fix the opponent’s utility \cite{press2012iterated}, extortion strategies, which guarantee that the player’s utility is no lower than the opponent’s \cite{press2012iterated,becks2019extortion}, and generous strategies, which promote cooperation and allow the opponent to obtain a higher utility \cite{stewart2013extortion}.

ZD strategies have gradually drawn increasing attention in CPS, while most existing studies focus on two-action games, and each player has only two available actions. In crowdsourcing systems, ZD strategies have been adopted to incentivize cooperation, where the requester and the worker repeatedly choose between cooperation and defection \cite{hu2019solving}. In audit and signaling games, ZD strategies have been employed by the defender through signaling and auditing, where the attacker chooses to attack or quit \cite{chen2022strategic}. In mining pool management, ZD alliances have been adopted in multi-player interactions, where each pool chooses between attack and non-attack actions \cite{tang2024cooperation}. Similarly, in blockchain systems, ZD strategies have been designed as incentive mechanisms for transaction trading \cite{feng2025zero}. 

Such two-action formulations typically correspond to scenarios where players face only two feasible targets. Only a few studies in CPS consider the multi-action games. Specifically, equalizer ZD strategies have been employed in IoT systems with multiple services \cite{wang2019moving}. Recall that equalizer ZD strategies represent only a restricted subset of the general ZD strategy space and are insufficient to fully exploit the diverse strategic interactions. Thus, the application of general ZD strategies to MTD problems with multiple targets remains challenging.

\textbf{Organization:}
Section II presents the security game model for MTD and SSE.  Section III shows the definition of ZD strategies and foundational properties. Section IV establishes programs for finding the ideal and the optimal ZD strategies. Section V computes the proposed ZD strategy and compares their complexity. Section VI gives experiments to evaluate our results, followed by the conclusions in Section VII.

\vspace{-5pt}

\section{Preliminary and formulation}

In this section, we present the security game model for MTD, revisit the leader-follower framework with SSE, and outline the problem in this paper.

\vspace{-10pt}

\subsection{Security game formulation}

The MTD strategy is an active defense technique in the repeated security game \cite{korzhyk2011stackelberg,li2022robust,wang2019moving,nguyen2019deception} between a defender $D$ and an attacker $A$ over the infinite time horizon $\mathbf{T}=\{0,1,\dots,t,\dots\}$. Take $\mathbf{P}=\{d,a\}$ as the player set. The attacker selects a target from $K$ ($K>1$) targets to invade, where the defender tries to prevent attacks by covering one target. Take $[K]=\{1,\dots,K\}$ as the target set. Let the action sets of the defender and the attacker be  $\mathcal{D}$ and  $\mathcal{A}$, respectively.  Naturally, we have $\mathcal{D}=\mathcal{A}=[K]$ in this security game setting, and each player chooses a target to protect or attack. Take $u_d: \mathcal{D}\times\mathcal{A} \to \mathbb{R}$ as the defender's utility function and $u_a:\mathcal{D}\times\mathcal{A} \to \mathbb{R}$ as the attacker's utility function.

Consider a widely studied repeated one-shot security game situation \cite{korzhyk2011stackelberg,ijcai2017-516,chen2009game}.
Let the coefficient $ U_d^c(k)$ be the defender’s profit if target $k$ is attacked while it is covered by the defender's protection. 
If target $k$ is uncovered by the defender,  the defender’s profit is depicted by the coefficient $U_d^u(k)$. 
Given an action profile $(d,a)\in\mathcal{D}\times\mathcal{A}$ of the defender and the attacker, take $\boldsymbol{x}=[x_1,\dots, x_K]^T$ as the defender's protection allocation over $K$ targets. Specifically,  $x_k=1$, if $d=k$, while $x_k=0$ if $d\neq k$. It implies the constraint $\sum_{k=1}^Kx_k=1$. The defender's one-shot utility function is
$
u_d(d,a)=x_{a} U_d^c(a)\!+\!(1-x_a) U_d^u(a).
$
Thus, the defender's utility $u_d$ is determined by the profit from the attacked target $a$, taking the value $U_d^c(a)$ if $x_a=1$ and $U_d^u(a)$ otherwise.
 Similarly, the attacker's one-shot utility function can be established with coefficients $U_a^c(k)$ and $ U_a^u(k)$, i.e.,
$
u_a(d,a)=x_a U_a^c(a)+(1-x_a) U_a^u(a).
$

In the repeated security game,  players usually choose memory-one strategies, where players' current actions depend on the outcomes from the previous stage \cite{nguyen2019deception}.
Let $d_{t}$ and $a_{t}$ denote the actions of the defender and the attacker at the current stage $t\geqslant 1$, respectively, whereas $d_{t-1}$ and $a_{t-1}$ represent their actions at the previous stage $t-1$. 
Take $\Delta\mathcal{D}$ as the set of probability distributions over the defender's action set $\mathcal{D}$.
The defender’s memory-one strategy is defined as a conditional probability distribution $\pi_d(\cdot|d_{t-1},a_{t-1})\in\Delta\mathcal{D}$. Specifically, $\pi_d(k|d_{t-1},a_{t-1})$ gives the probability that the defender chooses target $k$ at current stage, with the defender taking action $d_{t-1}$ and the attacker taking action $a_{t-1}$ at the previous stage.  Similarly, let $\Delta\mathcal{A}$ denote the set of probability distributions over the attacker's action set $\mathcal{A}$. The attacker's memory-one strategy is defined as $\pi_a(\cdot|d_{t-1},a_{t-1})\in\Delta\mathcal{A}$. Accordingly, the defender's current action, $d_t$, is drawn from $d_{t}\sim \pi_d(\cdot|d_{t-1},a_{t-1})$, and the attacker’s current action $a_t$ follows $a_{t}\sim \pi_a(\cdot|d_{t-1},a_{t-1})$.

In this setting, the repeated defender-attacker interactions induce dynamic utility changes. This utility changes and long-term interactions require both players to consider their cumulative utilities across all stages, no longer a single stage. We therefore model their objectives using the expected long-term utilities for both players \cite{feng2017stackelberg,li2022robust,cheng2023zero}: 
 \vspace{-5pt}
\begin{equation}\label{eq::utilty-longterm}
    \begin{aligned}
\bar{u}_d(\pi_d,\pi_a)&=\mathbb{E}\left(\lim_{T\to\infty}\sum\nolimits_{t=0}^T\frac{u_d(d_t,a_t)}{T}\right),\\
\bar{u}_a(\pi_d,\pi_a)&=\mathbb{E}\left(\lim_{T\to\infty}\sum\nolimits_{t=0}^T\frac{u_a(d_t,a_t)}{T}\right).
\end{aligned}
\end{equation}
Concisely, the repeated security game is denoted as $\mathcal{G}= \{\mathbf{P},\mathcal{D},\mathcal{A},\bar{u}_d,\bar{u}_a,\pi_d,\pi_a\}$. 
The defender seeks to maximize its expected long-term utility $\bar{u}_d$ through the memory-one strategy $\pi_d$. As it dynamically shifts protection across different targets over stages, $\pi_d$ is regarded as an MTD strategy.

\subsection{Revisiting SSE strategy}

In real-world security contexts, defensive strategies are often observable and persistent, remaining unchanged after deployment. This allows attackers to patiently observe and analyze the defender's strategies. Consequently, the interaction between the defender and the attacker is naturally modeled as a leader-follower
framework \cite{korzhyk2011stackelberg,vorobeychik2012computing}. Specifically, the defender is a leader and declares a strategy in advance, while the attacker is a follower and chooses its strategy after observing the defender's strategy. 

The attacker usually adopts a best response (BR) strategy to the defender's declared strategy $\pi_d$, as it is optimal to maximize the attacker's utility. Specifically, the set of the attacker's BR strategies is defined as:
 $ \textbf{{BR}}(\pi_d)=\mathop{\text{argmax}}_{\pi_a\in\Delta \mathcal{A}} \bar{u}_a(\pi_d,\pi_a).$
  Without loss of generality, the follower breaks ties optimally if there are multiple optimal strategies \cite{korzhyk2011stackelberg,9722864}. 
The defender aims to maximize
its expected utility considering the attacker, and the equilibrium is defined as 
 the strong Stackelberg equilibrium (SSE) \cite{korzhyk2011stackelberg,vorobeychik2012computing,9722864}.
\begin{definition}\label{def::SSE}
A strategy profile $(\pi_d^{SSE},\pi_a^{SSE})$ is said to be an SSE of the repeated security game $\mathcal{G}$ if
$$
\begin{aligned}
&(\pi_d^{SSE}, \pi_a^{SSE})\in\mathop{\text{argmax}}\limits_{\pi_d, \pi_a\in {\textbf{BR}}({\pi_d})}\bar{u}_d(\pi_d,\pi_a).
\end{aligned}
$$
\end{definition}


As SSE defines the defender's optimal strategy in the leader–follower framework, its computation has drawn significant interest. This problem is naturally addressed as a bi-level optimization: the defender selects $\pi_d^{SSE}$ at the upper level, while the lower level constraint ensures the attacker’s strategy lies in $\mathbf{BR}(\pi_d^{SSE})$. 
To solve the bi-level optimization problem, a common approach is to reformulate the attacker's optimization problem as a set of constraints \cite{korzhyk2011stackelberg,sinha2018stackelberg}.
Then the original bi-level optimization problem can be reformulated as a single-level optimization problem \cite{vorobeychik2012computing,9722864}. Let $Z$ denote a sufficiently large constant, and let $W, Q \in \mathbb{R}^{K \times K}$ be matrices representing the defender’s and attacker’s values associated with each state, respectively. 
Thus, the SSE can be calculated by a mixed-integer program as follows, whose proof is provided in Appendix \ref{app::le::1}.

\vspace{-5pt}
\begin{lemma}\label{le::compute_SSE}
The defender's SSE strategy $\pi_d$ can be solved by the following program \begin{equation}\label{eq::optimization::compute::SSE}
 \!\!  \begin{aligned}
\!\!\!\!\!\!\max\limits_{\substack{\pi_d,\pi_a,\\V_d,V_a,Q,W}} \! & V_d \\
\!\!\!\!\textnormal{s.t.} \
0&\!\geqslant \!\!\sum\nolimits_{d\!=\!1}^K \!\!\pi_d(d|i,j) \!\big(u_a(d,a) \!\!+ \!\!Q(d,a)\!\big)\!\!-\!\!V_a\!\! -\!\! Q(i,j) , \\
0&\!\!\leqslant\! \!\sum\nolimits_{d\!=\!1}^K \!\!\pi_d(d|i,j)\!\left(\!u_a(d,a)\! \!+\!\! Q(d,\!a)\!\right)\!\!-\!\! V_a\!\! - \!\!Q(i,j) \!\\
&\ \ + \!(1 - \pi_a(a|i,j))Z, \\
0 &\!\!\leqslant\!\! \sum\nolimits_{d\!=\!1}^K \!\!\pi_d(d|i,\!j)\!\left(\!u_d(d,\!k)\! \!+\! \!W(d,\!a)\!\right)\!\! -\!\! V_d \!\!-\!\! W(i,\!j)\\
&\ \ + (1 - \pi_a(a|i,j))Z,  \\
& \!\sum\nolimits_{k=1}^K \pi_a(k|i,j) = 1, \quad \pi_a(k|i,j) \in \{0,1\},\\
& \!\sum\nolimits_{k=1}^K \pi_d(k|i,j) = 1, \quad \pi_d(k|i,j) \geqslant 0. \\
\end{aligned}
\end{equation}
\end{lemma}

\vspace{-15pt}
\subsection{Problem statement}

Although many existing works solve the SSE computation problem via the mixed-integer program  (\ref{eq::optimization::compute::SSE}), it is worth noting that its computational complexity increases significantly when the target number $K$ is large.
 The complexity is driven by several factors. 
 First, the program involves a non-convex feasible region by addressing the attacker's BR strategy, which complicates the search for a globally optimal defender strategy~\cite{vorobeychik2012computing}.
Second, the presence of mixed-integer constraints further exacerbates the computational difficulty.
Even in mixed-integer linear programs, specialized algorithms like branch-and-bound often exhibit exponential time complexity under certain conditions \cite{basu2023complexity}.
Thus, the computation of the SSE may exhibit exponential complexity relative to the number of targets \cite{lopez2022stationary}, making the problem increasingly difficult to address in large-scale settings.

Generally, although the SSE strategy is optimal for the defender, the computational complexity imposes a heavy burden, preventing the defender from adopting effective MTD strategies in a timely manner. Therefore, this paper aims to address the following important problem:

\textbf{Problem 1:}
Find a new approach to construct the defender's MTD strategy with both the sufficiently high profit and significantly low complexity.

To solve Problem 1, we consider the following assumption, which is widely adopted in security problems \cite{korzhyk2011stackelberg,ijcai2019-75,ijcai2017-516}.

 \begin{assumption}\label{as::1}
  For $k\in[K]$, $U_d^c(k)> U_d^u(k)$.
\end{assumption}
Assumption \ref{as::1} makes sense for the defender's incentive to resist attacks, i.e., if target $k$ is attacked, the defender's utility of protecting target $k$ is higher than that of the unprotected target.

\section{ZD strategy}

As discussed above, the defender possesses a unilateral advantage by committing to a strategy where the attacker adopts a BR strategy. This naturally raises the question of whether Problem 1 can be addressed by exploiting such a unilateral advantage. ZD strategies have attracted increasing attention due to their ability to unilaterally enforce a self-determined linear relationship between the expected utilities of the players. In this section, we introduce the formal definition and foundational properties of ZD strategies in the security game $\mathcal{G}$.
 
\vspace{-10pt}

\subsection{Definitions of ZD strategy}

For  $k=1,\dots,K$, take \begin{equation}\label{eq::hat_pi_d}
    \hat{\pi}(k)=[\underbrace{\mathbf{0}^T_K,\dots, \mathbf{0}^T_K}_{k-1 \text { times }}, \mathbf{1}^T_K,\underbrace{\mathbf{0}^T_K,\dots, \mathbf{0}^T_K}_{K-k \text { times }}]^T,
\end{equation} and
$\pi_d(k)\!\!=\!\![\!\pi_d(k|1,\!1)\!,\!\dots\!,\!\pi_d(k|1,\!K)\!,\!\pi_d(k|2,\!K)\!,\!\dots\!\!,\!\pi_d(k|K\!,\!K)\!]^T\!\!\!,$
where $\mathbf{1}_K$ ($\mathbf{0}_K$) is a $K$-dimensional column vector with all elements of $1$ ($0$). Further, for $l\in\mathbf{P}$, let 
 $$S_l(k)=[U_l^u(1),\dots,U_l^u(k-1),U_l^c(k),U_l^u(k+1),\dots,U_l^u(K)]^T$$ and $\mathbf{S}^l=[S_l^T(1),\dots, S_l^T(K)]^T$ be the player's profit vector over all action profiles. 
In the following, we present the definition of the ZD strategy and explain why it unilaterally enforces a linear relationship between players' expected utilities.
\begin{figure*}[hb]
\vspace{-10pt}
\begin{equation}\label{eq::M}\begin{aligned}\!\!\!\!M(\pi_d,\pi_a)\!\!=\!\!\!\left[\begin{array}{llllll}
\pi_d(1|1,1)\pi_a(1|1,1) &\cdots &\pi_d(1|1,1)\pi_a(K|1,1) &\pi_d(2|1,1)\pi_a(1|1,1)&\cdots &\pi_d(K|1,1)\pi_a(K|1,1) \\
\pi_d(1|1,2)\pi_a(1|1,2)  &\cdots &\pi_d(1|1,2)\pi_a(K|1,2) &\pi_d(2|1,2)\pi_a(1|1,2)&\cdots &\pi_d(K|1,2)\pi_a(K|1,2) \\
\vdots
\\
\!\!\!\pi_d(1|K,\!K)\pi_a(1|K,\!K) \!\!\!\!\!&\cdots &\!\!\!\pi_d(1|K,K)\pi_a(K|K,K)\!\!\!\!\! &\!\!\pi_d(2|K,K)\pi_a(1|K,K)\!\!\!\!\!&\cdots &\!\!\pi_d(K|K,K)\pi_a(K|K,K)\!\! \!\!\!
\end{array}\right]
\end{aligned}\!.\end{equation}
\end{figure*}
\begin{definition}\label{def::ZD}
The defender's strategy $\pi_d$ is a ZD  strategy if 
 there exist  $\alpha$, $\beta$, $\gamma\in\mathbb{R}$, and  $\phi_k\in\mathbb{R}$, such that
\begin{equation}\label{eq::ZD-def-muliti}
\begin{aligned}
\sum\nolimits_{k=1}^{K}\phi_k\left(\pi_d(k)-\hat{\pi}(k)\right)&=\alpha \mathbf{S}^{d} +\beta \mathbf{S}^a +\gamma \mathbf{1}_{K^2},\\
\sum\nolimits_{k=1}^{K}\pi_d(k)&=\mathbf{1}_{K}.
\end{aligned}
\end{equation}
\end{definition}

For any defender's strategy $\pi_d$ and attacker's strategy $\pi_a$, the state transition matrix $M(\pi_d,\pi_a)$ is defined by (\ref{eq::M}), and the stationary vector $\mathbf{v}$ satisfies $\mathbf{v}^T(M(\pi_d,\pi_a)-I)=0$.
Let $\mathbf{f}=[f_1,f_2,\dots,f_{K^2}]^T$ be an arbitrary vector. Define the diagonal matrices  $E_1=\text{diag}(\underbrace{1,\dots, 1}_{K^2-1 \text { times }},0)$, $E_2=\text{diag}(\underbrace{0,\dots, 0}_{K^2-1 \text { times }},1)$,  and construct the modified matrix $D(\pi_d,\pi_a,\mathbf{f})=(M(\pi_d,\pi_a)-I)E_1+\mathbf{f}\mathbf{1}_{K^2}^TE_2$. Notice that $D(\pi_d,\pi_a,\mathbf{f})$ replaces the last column of  $M(\pi_d, \pi_a) - I$ with $\mathbf{f}$. Then, as shown in \cite{press2012iterated,tan2021payoff},  $$\mathbf{v}^T \mathbf{f} =det(D(\pi_d,\pi_a,\mathbf{f})).$$
Using this identity, the expected utilities of the defender and the attacker in (\ref{eq::utilty-longterm}) can be rewritten as 
\begin{equation}\label{eq::expectuti}
    \begin{aligned}
        \bar{u}_d(\pi_d,\pi_a)=\frac{det(D(\pi_d,\pi_a,\mathbf{S}^d))}{det(D(\pi_d,\pi_a,\mathbf{1}_{K^2}))}, \\
    \bar{u}_a(\pi_d,\pi_a)=\frac{det(D(\pi_d,\pi_a,S^a))}{det(D(\pi_d,\pi_a,\mathbf{1}_{K^2}))}.
    \end{aligned}
\end{equation} 
Thus, for any parameters $\alpha$, $\beta$, and $\gamma\in\mathbb{R}$, we have
\begin{equation}
\begin{aligned}
    \alpha\bar{u}_d(\pi_d,\pi_a)+\beta \bar{u}_a(&\pi_d,\pi_a)+\gamma\\
    =&\frac{det(D(\pi_d,\pi_a,\alpha S^d+\beta S^a+\gamma))}{det(D(\pi_d,\pi_a,\mathbf{1}_{K^2}))}.
\end{aligned}
\end{equation}
Besides, for $k\in[K-1]$, $\pi_d(k)-\hat{\pi}(k)$ is the sum of the $K$ vectors from the $\left((k-1)K+1\right)$-th  column to $\left(kK\right)$-th column of $D(\pi_d,\pi_a,\alpha S^d+\beta S^a+\gamma)$. Thus, if the defender's strategy $\pi_d$ satisfies (\ref{eq::ZD-def-muliti}), then the last column of $D(\pi_d,\pi_a,\alpha S^d+\beta S^a+\gamma)$ is a linear combination of the first $(K-1)K$ columns, that is, $det(D(\pi_d,\pi_a,\alpha S^d+\beta S^a+\gamma))=0$. Therefore, once the defender determines the linear parameters $\alpha$, $\beta$, $\gamma$, and chooses a ZD strategy $\pi_d$ according to these parameters in Definition \ref{def::ZD}, the expected utilities of the two players satisfy the following linear relationship:
\begin{equation}\label{eq::def::ZD}\alpha \bar{u}_d(\pi_d,\pi_a)+\beta \bar{u}_a(\pi_d,\pi_a)+\gamma=0, \ \forall  \pi_a\in \Delta\mathcal{A}.\end{equation}
Accordingly, we refer to $\alpha$, $\beta$, and $\gamma$ as the linear parameters of the ZD strategy. Further, as shown in Definition \ref{def::ZD}, given any linear parameters, the parameters $\{\phi_k\}_{k=1}^K$ determine the feasibility of ZD strategies, and we thereby term $\{\phi_k\}_{k=1}^K$  the feasibility parameters.

Since a player adopting a ZD strategy can unilaterally determine the linear relationship between the players’ expected utilities, the linear parameters $\alpha$, $\beta$, and $\gamma$ play a crucial role. Depending on their configurations, several representative typical cases of ZD strategies can be constructed for different strategic purposes.
We introduce three typical ZD strategies:

(1) \textit{Equalizer strategy}: By taking $\alpha=0$ in Definition \ref{def::ZD}, we obtain the equalizer strategy, which satisfies  $\sum_{k=1}^{K}\phi_k\left(\pi_d(k)-\hat{\pi}(k)\right)=\beta \mathbf{S}^{a}+\gamma \mathbf{1}_{K^2}$. The player with an equalizer strategy can equalize the opponent's utility, no matter what strategy the opponent selects  \cite{press2012iterated}.

(2) \textit{Extortion strategy}: Let $\beta=-\chi\alpha$ with an extortion factor $\chi\geqslant 1$. The corresponding  extortion strategy fulfills $\sum_{k=1}^{K}\phi_k\left(\pi_d(k)-\hat{\pi}(k)\right)=\phi[(\mathbf{S}^{d}-\theta\mathbf{1})-\chi(\mathbf{S}^{a}-\theta\mathbf{1})]+\mathbf{p}_0$. 
With $\chi\geqslant 1$, the player who adopts extortion strategies can ensure that any improvement in its own utility exceeds that of the opponent \cite{press2012iterated,becks2019extortion}.

(3) \textit{Generous strategy}:  A generous strategy is defined by $\beta=-\chi\alpha$ with a generous factor $\chi\leqslant 1$, and takes the form: $\sum_{k=1}^{K}\phi_k\left(\pi_d(k)-\hat{\pi}(k)\right)=\phi[(\mathbf{S}^{d}-\theta\mathbf{1})-\chi(\mathbf{S}^{a}-\theta\mathbf{1})]+\mathbf{p}_0$. The utility improvement of the player adopting a generous strategy is not higher than that of the opponent, thereby promoting compromise and cooperation \cite{stewart2013extortion}.

\begin{remark}
    Most existing studies on CPS have mainly focused on ZD strategies in two-target settings \cite{hu2019solving,chen2022strategic,tang2024cooperation,feng2025zero,cheng2023zero}. 
 In such cases with two targets, the first equality in (\ref{eq::ZD-def-muliti}) lies in $2^2$-dimensional vectors, which permits a tractable analysis of specific ZD properties. In multi-target scenarios, however, this dimension increases to $K^2$, and the strategy form becomes substantially complex as $K$ grows. As a result, in multi-target scenarios, the properties of general ZD strategies appear to be much more intricate than those in the two-target case.
\end{remark}

\vspace{-5pt}
\subsection{Foundational properties of ZD}

With the definition of the ZD strategy, we are naturally concerned with its existence and performance.
In the context of security games, the linear parameters, $\alpha$, $\beta$, and $\gamma$, of the defender’s ZD strategy must belong to a feasible set, which ensures the existence of ZD strategies $\pi_d^{ZD}$. 
Take 
$$\begin{aligned}
    \phi^{\max}&=\max\{\phi_1,\dots,\phi_{K}\},\ 
    \phi^{\min}=\min\{\phi_1,\dots,\phi_{K}\}.\\
\end{aligned}$$
We define $\phi^{\max}_{-k} = \max\{\phi_1, \dots, \phi_{k-1}, \phi_{k+1}, \dots, \phi_K\}$ as the maximum value except $\phi_k$, and similarly define $\phi^{\min}_{-k}$ as the minimum value.
The following theorem provides a sufficient and necessary condition for the existence of ZD strategies, whose proof can be found in Appendix \ref{app::th::1}.

\begin{theorem}[Existence of ZD strategy]\label{th::1}
For any linear parameters $\alpha$, $\beta$, and $\gamma$, there exists a ZD strategy $\pi_d^{ZD}$ enforcing  $\alpha \bar{u}_d(\pi_d^{ZD},\pi_a)+\beta \bar{u}_a(\pi_d^{ZD},\pi_a)+\gamma=0$, if and only if there exist $\phi_1,\dots,\phi_{K-1}\geqslant0$ and $\phi_K=0$  such that,  for $k\in[K]$,  
\begin{equation}\label{eq::ZD-parameter}
\begin{aligned}
-\phi_k&\leqslant\alpha  U_d^c(k)+\beta U_a^c(k)+\gamma\leqslant \phi^{\max}-\phi_k,\\
-\phi^{\min}_{-k}&\leqslant\alpha  U_d^u(k)+\beta U_a^u(k)+\gamma\leqslant\phi^{\max}-\phi^{\max}_{-k}.
\end{aligned}
\end{equation}
\end{theorem}

Theorem \ref{th::1} shows that the ZD strategy with the corresponding linear relation exists, if the linear parameters satisfy inequalities (\ref{eq::ZD-parameter}). The first line in (\ref{eq::ZD-parameter}) indicates a utility's relation when the defender protects the target successfully, while the second line in (\ref{eq::ZD-parameter}) shows the other relation when the attacker invades the target successfully. These two different cases are important features of security games, which are distinguished from zero-sum and symmetric games. 

With the ZD existence, it is necessary to further examine its performance. According to Definition~\ref{def::SSE}, the SSE strategy always provides the highest utility for the defender when facing an attacker adopting the BR strategy. Therefore, evaluating the upper-bound performance of a ZD strategy is important, and obviously, the SSE strategy serves as an appropriate baseline for comparison, as proved in Appendix \ref{app::th::2}.

\vspace{-5pt}
\begin{theorem}[Upper-bound performance of ZD strategy]\label{le::ZDlemma2}
    Given Assumption \ref{as::1} and the defender's ZD strategy $\pi_d^{ZD}$, 
    \begin{equation}\label{eq::th2222::ideal performance}
    \bar{u}_d(\pi_d^{ZD},\textbf{{BR}}(\pi_d^{ZD}))\leqslant \bar{u}_d(\pi_d^{SSE},\pi_a^{SSE}),
    \end{equation}
    where $(\pi_d^{SSE},\pi_a^{SSE})$ is the SSE of the security game $\mathcal{G}$.
\end{theorem}

\begin{figure}[tbp]
\centering
 \includegraphics[width=2.4in]{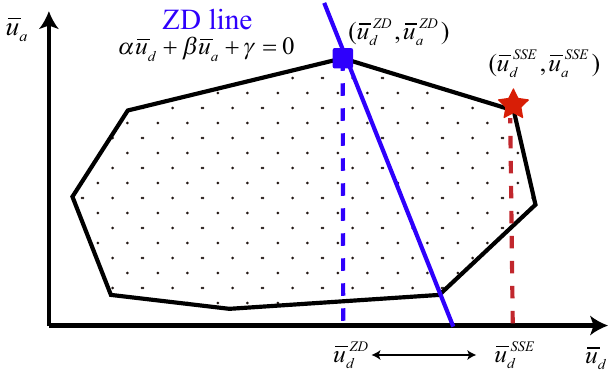}
 \vspace{-10pt}
\caption{Schematic comparison of the players' expected utility pairs $(\bar{u}_d,\bar{u}_a)$ under the SSE and ZD strategies. 
 The red star indicates the SSE outcome, corresponding to   $(\bar{u}_d^{SSE},\bar{u}_a^{SSE})=\left(\bar{u}_d(\pi_d^{SSE},\pi_a^{SSE}),\bar{u}_a(\pi_d^{SSE},\pi_a^{SSE})\right)$,  while the blue square denotes the ZD outcome, corresponding to $(\bar{u}_d^{ZD},\bar{u}_a^{ZD})=\left(\bar{u}_d(\pi_d^{ZD},\textbf{{BR}}(\pi_d^{ZD})),\bar{u}_a(\pi_d^{ZD},\textbf{{BR}}(\pi_d^{ZD}))\right)$. The blue line represents the ZD line $\alpha\bar{u}_d+\beta \bar{u}_a+\gamma=0$. The figure illustrates a performance gap for the defender, i.e., $\bar{u}_d^{SSE}>\bar{u}_d^{ZD}$.
 }
\label{fi::SSEvsZD}
 \vspace{-13pt}
\end{figure}

In particular, when a ZD strategy achieves the upper-bound performance for the defender, that is, the equality in (\ref{eq::th2222::ideal performance}) holds, we refer to the ZD strategy as the \textbf{ideal ZD strategy}. However, such an ideal strategy may not always exist.  As illustrated in  Fig.~\ref{fi::SSEvsZD}, 
due to the feasible utility pairs and the constraints on ZD parameters, it is not always possible to construct a ZD linear relationship that passes through the SSE outcome, which results in a performance gap between the SSE and ZD strategies for the defender. In these cases, it is reasonable to adopt the \textbf{optimal ZD strategy}, which yields the optimal performance for the defender among all feasible ZD strategies.
Since Theorem \ref{th::1} guarantees the existence of ZD strategies and Theorem \ref{le::ZDlemma2} establishes the upper-bound performance of the ZD strategy, we investigate how to select the ZD strategy in the following section.

\section{Program for ZD identification}

Note that any ZD strategy is characterized by the parameters $\alpha$, $\beta$, and $\gamma$. To determine a specific ZD strategy, we establish how these parameters should be constructed in this section.

\subsection{For ideal ZD strategy}

Since the SSE strategy serves as the optimal defense strategy for the defender in Theorem~\ref{le::ZDlemma2}, it is natural to examine whether there exists an ideal ZD strategy that can provide the defender with the best profit. The following theorem establishes the condition for the existence of such an ideal ZD strategy, the proof of which is presented in Appendix \ref{app::th::3}.

\begin{theorem}[Program for the ideal ZD]\label{th::2}
Under Assumption \ref{as::1}, the defender's ZD strategy can bring the best profit 
if linear parameters $\alpha,\beta,\gamma$ are feasible solutions of  the program:
%
\begin{equation}\label{eq::th::2}
\begin{aligned}
\max\limits_{\alpha,\beta,\gamma}\ &0\\
\textnormal{s.t.} \ &0= \alpha U_d^c (1)+\beta U_a^c (1)+\gamma,\\
&0\leqslant \alpha U_d^c (K)+\beta U_a^c (K)+\gamma ,\\
&0\leqslant\alpha U_d^u (1)+\beta U_a^u (1)+\gamma,\\
&0=\alpha U_d^u (k)+\beta U_a^u (k)+\gamma , ~k=2,\dots,K-1,\\
&0\geqslant\alpha U_d^u (K)+\beta U_a^u (K)+\gamma,\\
 &0 \geqslant \alpha, 0\leqslant  \beta.\\
\end{aligned}
\end{equation}
\end{theorem}

When $\alpha$, $\beta$, and $\gamma$ satisfy the condition in program (\ref{eq::th::2}), the defender can achieve the same best profit as that with the SSE strategy stated in Theorem \ref{le::ZDlemma2}, which confirms the existence of an ideal ZD strategy. Besides, Theorem \ref{th::2} offers a practical method for constructing such a strategy: by solving program (\ref{eq::th::2}) for $\alpha,\beta$, and $\gamma$, one can directly derive the linear parameters for an ideal ZD strategy.

Moreover, it is notable that the constraints in program (\ref{eq::th::2}) consist of only $K+4$ linear inequalities and equations. In contrast to the non‑convex constraints and mixed‑integer variables in the program (\ref{eq::optimization::compute::SSE}), program (\ref{eq::th::2}) delivers lower
computational complexity while offering the same profit as the SSE strategy. Thus, when the constraints in program (\ref{eq::th::2}) are satisfied, the defender can employ such an ideal ZD strategy as a computationally efficient alternative MTD strategy.

Focusing on typical cases of ZD strategies discussed in the previous section, we can derive simplified versions of program (\ref{eq::th::2}), each tailored to address different defense requirements.

\begin{corollary}[Equalizer ZD]\label{co::1}
Under Assumption \ref{as::1}, if $U_a^c(K)\geqslant U_a^c(1)$, $U_a^u(1)\geqslant U_a^c(1)$, $U_a^u(K)\leqslant U_a^c(1)$, and $U_a^u(k)= U_a^c(1)$, for $k=2,\dots,K$, then a defender's equalizer ZD strategy can bring the best profit.
\end{corollary}

\begin{corollary}[Extortion ZD]\label{co::2}
Under Assumption \ref{as::1}, if
$\chi=\frac{U_d^c (1)-\theta}{U_a^c (1)\!-\!\theta}\geqslant 1$,
$
0\leqslant (U_d^c (K)\!-\!\theta)-\chi( U_a^c (K)\!-\!\theta)$, $
0\geqslant (U_d^u (K)\!-\!\theta)-\chi( U_a^u (K)\!-\!\theta)$, $
 0\leqslant (U_d^u (1)\!-\!\theta)-\chi( U_a^u (1)\!-\!\theta)$, and $
0=(U_d^u (k)\!-\!\theta)-\chi( U_a^u (k)\!-\!\theta)$, for $k=2,\dots,K$,
then a defender's extortion ZD strategy can bring the best profit.  
\end{corollary}

\begin{corollary}[Generous ZD]\label{co::3}
Under Assumption \ref{as::1}, if 
$\chi=\frac{U_d^c (1)-\theta}{U_a^c (1)\!-\!\theta}\leqslant 1$,
$
0\leqslant (U_d^c (K)\!-\!\theta)-\chi( U_a^c (K)\!-\!\theta)$, $
0\geqslant (U_d^u (K)\!-\!\theta)-\chi( U_a^u (K)\!-\!\theta)$, $
 0\leqslant (U_d^u (1)\!-\!\theta)-\chi( U_a^u (1)\!-\!\theta)$, and $
0=(U_d^u (k)\!-\!\theta)-\chi( U_a^u (k)\!-\!\theta)$, for $k=2,\dots,K$,
then a defender's generous ZD strategy can bring the best profit.  
\end{corollary}


\subsection{For optimal ZD strategy}

Since such an ideal ZD strategy may not always exist, it is reasonable to employ an optimal ZD strategy to mitigate the defender’s utility loss when deviating from the SSE strategy. In the following, we consider that the defender, as the leader, aims to obtain an optimal strategy in the ZD strategy set to maintain a good defensive performance and avoid adopting the SSE strategy, which brings a high computational cost.

To obtain the optimal ZD strategy, we introduce the following notations. Take $$ \!\Lambda(i_1,i_2)\!\!=\!\! \left\{\!\!(\alpha,\beta,\gamma)\!\left| \begin{aligned}
&\alpha  U_d^c(i_1)+\beta U_a^c(i_1)+\gamma\leqslant 0,\\
&\alpha  U_d^u(i_1)+\beta U_a^u(i_1)+\gamma \geqslant 0,\\
&\alpha  U_d^c(i_2)+\beta U_a^c(i_2)+\gamma\geqslant 0,\\
&\alpha  U_d^u(i_2)+\beta U_a^u(i_2)+\gamma\leqslant0,\\
&\alpha  U_d^u(k)+\beta U_a^u(k)+\gamma \!=\!0, k \!\neq \!i_1,i_2.
\end{aligned}\right.\!\!\!\right\}\!,$$ and define the set $\Lambda$ as the union of all distinct $\Lambda(i_1,i_2)$, i.e., $\Lambda=\cup_{i_1\neq i_2}\Lambda(i_1,i_2)$, to denote all possible linear parameter sets. In addition, we define $$\text{Conv}(G)=\text{span}\{(U_d^c(k), U_a^c(k)),(U_d^u(k),U_a^u(k)), k \in[K]\}$$ as the convex hull of all possible utility pairs. 
In the following, we formulate solving a program to determine the parameters $\alpha$, $\beta$, and $\gamma$ corresponding to the optimal ZD strategy, with the proof in Appendix \ref{app::th::4}.

\begin{theorem}[Program for the optimal ZD]\label{th::optimal_ZD}

Under Assumption \ref{as::1}, one can find the linear parameters $\alpha,\beta,$ and $\gamma$ regarding the defender's optimal ZD strategy by solving the following program:
\begin{equation}\label{eq::optimization 2}
    \begin{aligned}
\max\limits_{\alpha,\beta,\gamma,u_d,u_a}  &u_d\\
\textnormal{s.t.}  \ \ &(\alpha,\beta,\gamma) \in \Lambda, \\
& (u_d,u_a)\in \text{Conv}(G),\\
& \alpha u_d+\beta u_a + \gamma =0. 
\end{aligned}
\end{equation}

\end{theorem}

We begin by providing a detailed explanation for each constraint in (\ref{eq::optimization 2}),  individually.
 $(\alpha,\beta,\gamma) \in \Lambda$ examines the set of all linear parameters of feasible ZD strategies, and $ (u_d,u_a)\in \text{Conv}(G)$ explores all possible players' utility pairs, while $\alpha u_d+\beta u_a + \gamma =0$ demonstrates the relation between players' utilities and the linear parameters. 
 Although the program (\ref{eq::optimization 2}) consists of many inequalities and equations, it remains tractable due to the fixed number of variables, which is limited to five. In contrast to program (\ref{eq::optimization::compute::SSE}), the reduction of variables makes program (\ref{eq::optimization 2}) much simpler and solvable.

Both Theorem \ref{th::2} and Theorem \ref{th::optimal_ZD} characterize how to construct the linear parameters $\alpha$, $\beta$, and $\gamma$ for the corresponding ZD strategy, thereby forming the foundation for the algorithm design of the strategy $\pi_d$. The key difference between the two theorems lies in their solution methodologies. 
Theorem \ref{th::optimal_ZD} provides a more general approach than Theorem \ref{th::2} for identifying linear parameters that yield a ZD strategy approximating the utility of SSE strategies. From a computational perspective, solving (\ref{eq::optimization 2}) in Theorem \ref{th::optimal_ZD} is generally more complex than solving (\ref{eq::th::2}) in Theorem \ref{th::2}. Nevertheless, both of them are significantly more tractable than computing the SSE strategy directly. Consequently, these two theorems offer practical and efficient approaches for computing the ZD strategy, where the detailed algorithm design and complexity analysis will be presented in the next section.

\section{ZD strategy computation}

In the previous section, we have shown how to find the linear parameters corresponding to the ZD strategy. Based on these results, we now present the concrete computation of the ZD strategy as well as analyze the computational complexity.

\vspace{-5pt}

\subsection{Algorithm design}

As shown in Definition \ref{def::ZD}, the defender's ZD strategy $\pi_d$ depends on the linear parameters $\alpha$, $\beta$, $\gamma$, and the feasibility parameters $\{\phi_k\}_{k=1}^K$. It inspires us to compute the ZD strategy in the following three steps.
\begin{itemize}
    \item \textbf{To find linear parameters $\alpha$, $\beta$, and $\gamma$:} The linear parameters $\alpha$, $\beta$, and $\gamma$ of the ZD strategy need to be selected to yield a satisfactory utility for the defender. According to Theorem \ref{th::2}, if an ideal ZD strategy exists, these parameters can be determined by solving program (\ref{eq::th::2}) for the upper-bound performance. Otherwise, the optimal ZD strategy can be obtained by solving the program (\ref{eq::optimization 2}) in Theorem \ref{th::optimal_ZD} to maximize the defender's utility.

 \item \textbf{To construct feasibility parameters $\{\phi_k\}_{k=1}^K$:}  
Since the conditions in (\ref{eq::ZD-parameter}) admit multiple feasible solutions for the parameters $\{\phi_k\}_{k=1}^K$, we only need to present one explicit construction. We first consider a series of  $\{\phi_k\}_{k=1}^K$ with $\phi_K$, $\phi_{K-1}$, $\phi_1$ as the minimum, second minimum, and the maximum values, respectively.  By enforcing this ordering with (\ref{eq::ZD-parameter}), we sequentially obtain a feasible  $\{\phi_k\}_{k=1}^K$.

    \item \textbf{To compute the strategy $\pi_d$:}
Given the feasibility parameters, the strategy $\pi_d$ could be computed sequentially by taking $\hat{\pi}(k)$ in (\ref{eq::hat_pi_d}). Specifically, at step $k$, the residual term $\alpha \mathbf{S}^{d} \!+\!\beta \mathbf{S}^a \!+\!\gamma \mathbf{1}_{K^2}\!-\!\sum_{i=k+1}^{K-1}\!\phi_i\left(\pi_d(i)\!-\!\hat{\pi}(i)\right)$ is treated as a fixed quantity. Then, the weighted parameters $\{\omega_k\}_{k=1}^K$ are applied to obtain a valid realization of $\pi_d$ that satisfies the constraint in (\ref{eq::ZD-def-muliti}). 

\end{itemize}
The overall computation of the defender's ZD strategy is summarized as in Algorithm \ref{Al::ZD construction}. Besides, we introduce Theorem \ref{pro::algorithmZD} to 
show the optimality guarantee for the output of Algorithm \ref{Al::ZD construction}, whose proof is shown in Appendix \ref{app::th::5}.

	\begin{algorithm}[t]
 
		\caption{ZD strategy computation}
  \label{Al::ZD construction}
		\begin{algorithmic}[1]
			\REQUIRE \! \!\! \!\!Configurations $U_l^c\!(\!k\!)$ and $U_l^u\!(\!k\!)$, for $\!l\!\in\!\mathbf{\!P\!}\!,k \!\in\![K]$;\\
           \quad  \quad \quad\quad \ Weighted parameters $\{\!\omega_k\!\}_{k\!=\!1}^K$ with $\sum_{k\!=\!1}^K \!\omega_k \!= \!1$.
            \STATE \textbf{Find the linear parameters, $\alpha$, $\beta$, and $\gamma$:}\\
            \textbf{if} there exist a feasible  solution to (\ref{eq::th::2}), \textbf{then} \\
            \quad return $\alpha$, $\beta$, and $\gamma$.\\
            
            \textbf{else} \\
            \quad address (\ref{eq::optimization 2}) and return the optimal  solution $\alpha$, $\beta$, $\gamma$.\\
            \textbf{end if}
            \STATE \textbf{Construct feasibility parameters $\phi_1,\dots,\phi_{K}$:}\\
 $\phi_K=0.$\\
 $\begin{aligned}\phi_{K-1} & = \max\left\{|\alpha U_d^c (K)+\beta U_a^c (K)+\gamma|, \right.\\
&\quad\quad\quad\ \  \left.| \alpha U_d^u(K)+\beta U_a^u(K)+\gamma|\right\}.\end{aligned}$\\
\textbf{for} $k=K-2,\dots,2$ \textbf{do}\\
 \quad \ \ $\phi_k=|\alpha U_d^c (k)+\beta U_a^c (k)+\gamma|+\phi_{K-1},$\\
\textbf{end for}\\
 $\begin{aligned}
     \phi_1= &2\sum\limits_{k=2}^{K-1}\phi_k+|\alpha U_d^u (1)+\beta U_a^u (1)+\gamma |\\
     &+|\alpha U_d^c (1)+\beta U_a^c (1)+\gamma |.
 \end{aligned}$\\
return $\phi_1,\dots,\phi_{K}$.\\
\STATE \textbf{Compute defender's ZD strategy $\pi_d$:}\\
\textbf{for} $k=K-1,K-2,\dots,1$ \textbf{do}\\
\quad \ \ $\pi_d (k)\!=\!\left[\!\frac{\alpha \mathbf{S}^{d} \!+\!\beta \mathbf{S}^a \!+\!\gamma \mathbf{1}_{K^2}\!-\!\sum\limits_{i=k+1}^{K-1}\!\phi_i\left(\pi_d(i)\!-\!\hat{\pi}(i)\right)\!-\!\omega_k}{\phi_k}\!+\!\hat{\pi}(k)\right]^+\!\!.$\\
\textbf{end for}

 $\pi_d(K)=1-\sum\limits_{k=1}^{K-1}\pi_d(k).$\\
 Return $\pi_d$.

		\end{algorithmic}
	\end{algorithm}



 \begin{theorem}[Optimality guarantee]\label{pro::algorithmZD}
Given the game configuration $U_d^c(k)$, $U_d^u(k)$, $U_a^u(k)$, $U_d^a(k)$ for $k \in[K]$, the output strategy $\pi_d$ of Algorithm \ref{Al::ZD construction} yields the optimal utility among all the defender’s feasible ZD  strategies.

 \end{theorem}

\subsection{Computational Complexity}

Next, we analyze the computational complexity with the SSE strategy and the ZD strategy through Algorithm \ref{Al::ZD construction}.

\textbf{Complexity of computing SSE:}
The mixed-integer program in (\ref{eq::optimization::compute::SSE}) is NP-hard \cite{basu2023complexity}, implying that no deterministic polynomial-time exact algorithm exists unless P $=$ NP. The computational challenge stems from two main sources: the non-convexity of constraints and the presence of binary variables $\pi_a(k|i,j)\in\{0,1\}$.
For convex mixed-integer polynomial programs, branch-and-bound algorithms exhibit worst-case complexity exponential in the number of integer variables \cite{basu2023complexity,basu2025information}. Our formulation, involving $K^3$ binary variables, leads to a worst-case time complexity of $O(2^{K^3})$. Even in the simpler case of mixed-integer linear programs, computing an exact solution remains computationally demanding. Existing theoretical analyses of branch-and-bound methods, possibly augmented with cutting-plane techniques, yield complexity bounds such as $O((M')^{2K})$ under specific assumptions~\cite{basu2023complexityBB}, which are still exponential in $K$.
This complexity is inherent to the underlying problem of computing SSE in repeated games, not merely a consequence of our single-level mixed-integer programming reformulation. The inherent difficulty is further supported by complexity theory: finding Nash equilibria in general-sum stochastic games is PPAD-complete \cite{deng2023complexity, chen2022finding}, which strongly suggests that no polynomial-time exact algorithm exists for the related equilibrium problem.


\begin{figure*}[ht]
\centering
 \vspace{-10pt}
 \includegraphics[width=6.2in]{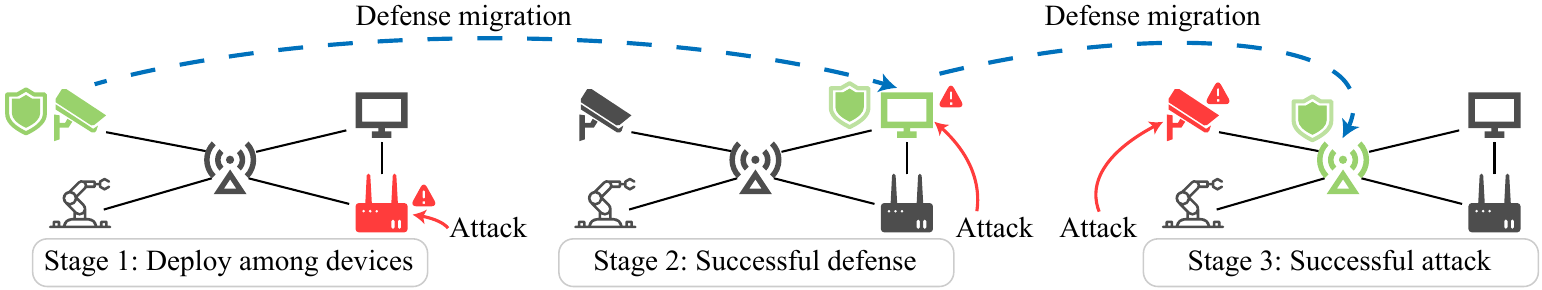}
 \vspace{-5pt}
\caption{MTD strategy in an IoT system
}
\label{fi::Iot}
\vspace{-5pt}
\end{figure*}
\begin{figure*}[ht]
\vspace{-5pt}
\centering
\subfigure[$K=3,\zeta=1$]{\label{fi::1a}
\begin{minipage}[t]{0.32\linewidth}
\centering
\includegraphics[width=2in]{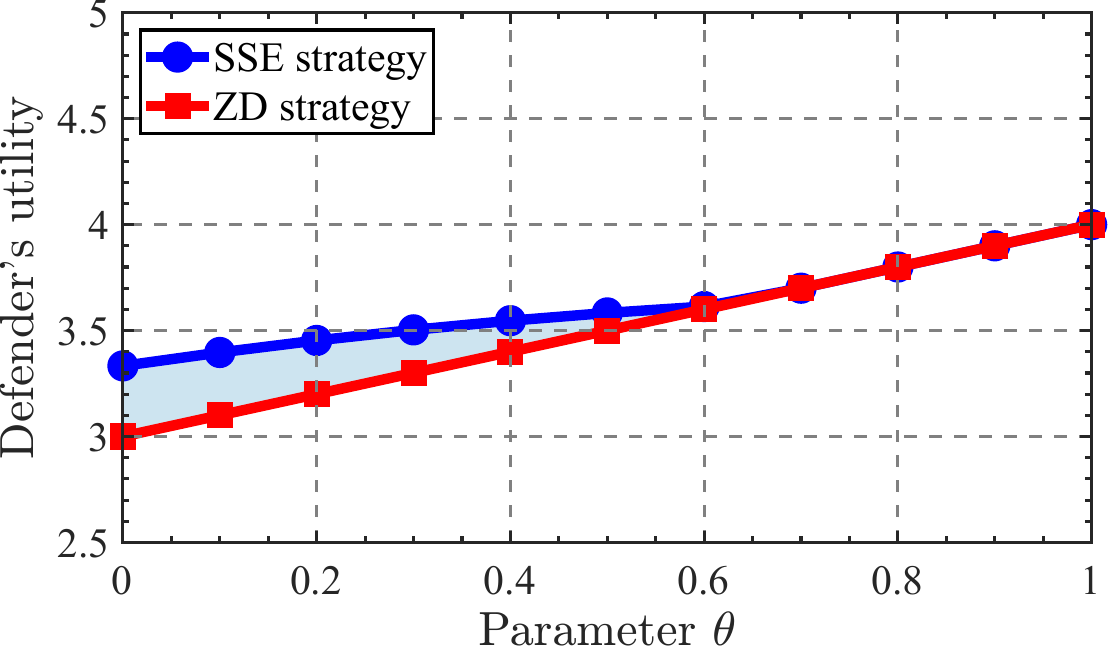}
\end{minipage}%
}%
\subfigure[$K=3,\zeta=2$]{\label{fi::1b}
\begin{minipage}[t]{0.32\linewidth}
\centering
\includegraphics[width=2in]{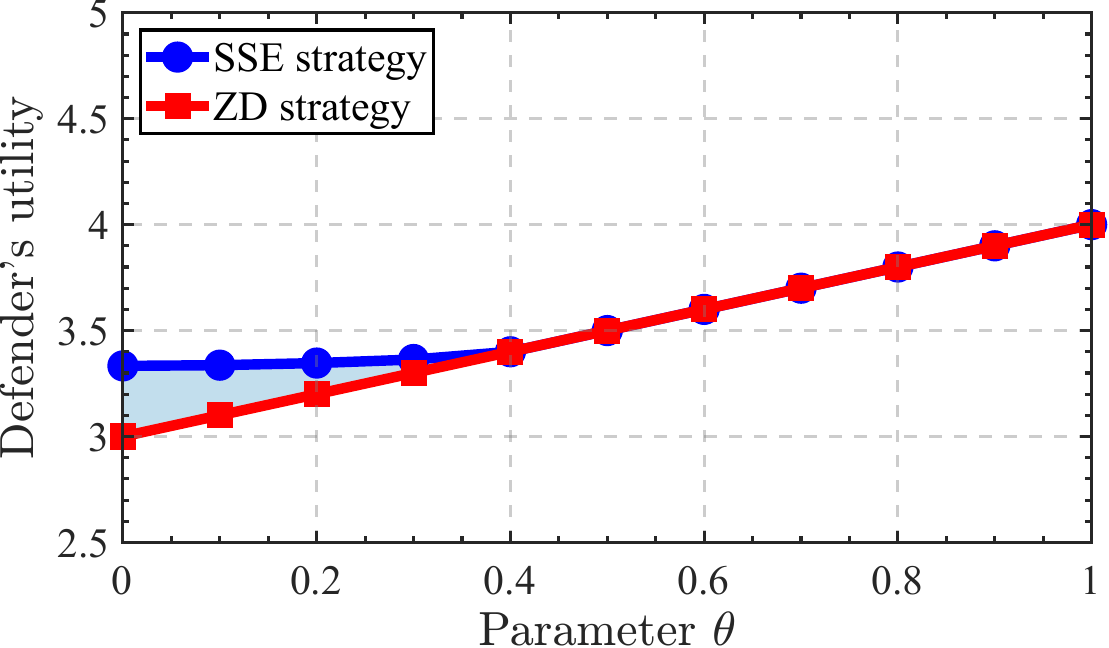}
\end{minipage}%
}%
\subfigure[$K=3,\zeta=3$]{\label{fi::1c}
\begin{minipage}[t]{0.32\linewidth}
\centering
\includegraphics[width=2in]{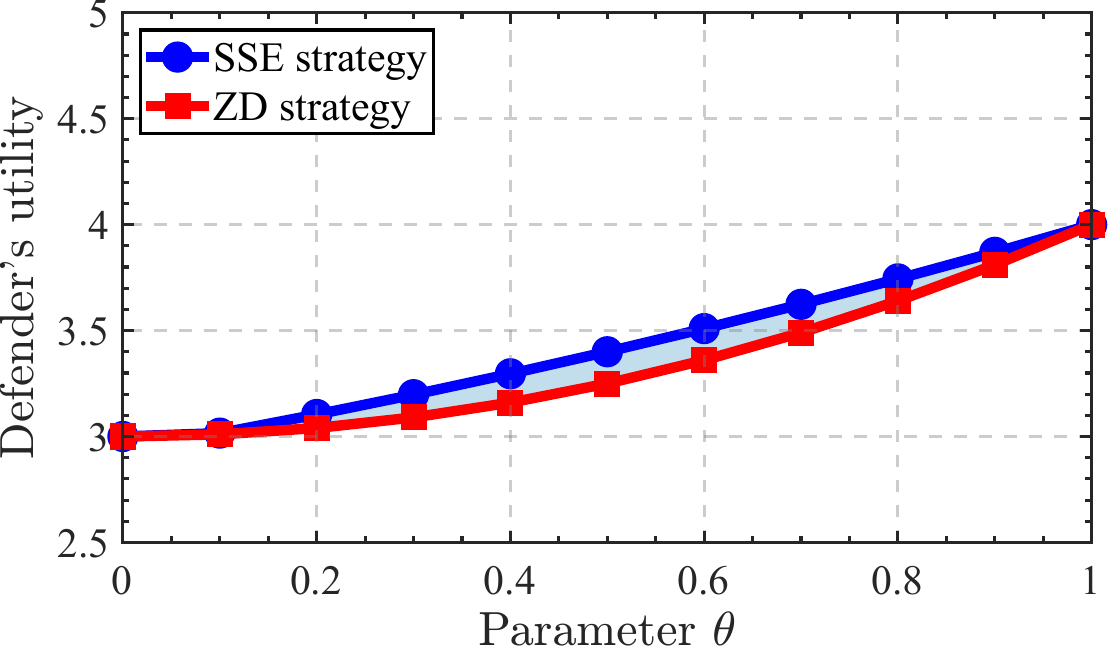}
\end{minipage}%
}\\
\vspace{-5pt}
\subfigure[$K=5,\zeta=1$]{\label{fi::1a}
\begin{minipage}[t]{0.32\linewidth}
\centering
\includegraphics[width=2in]{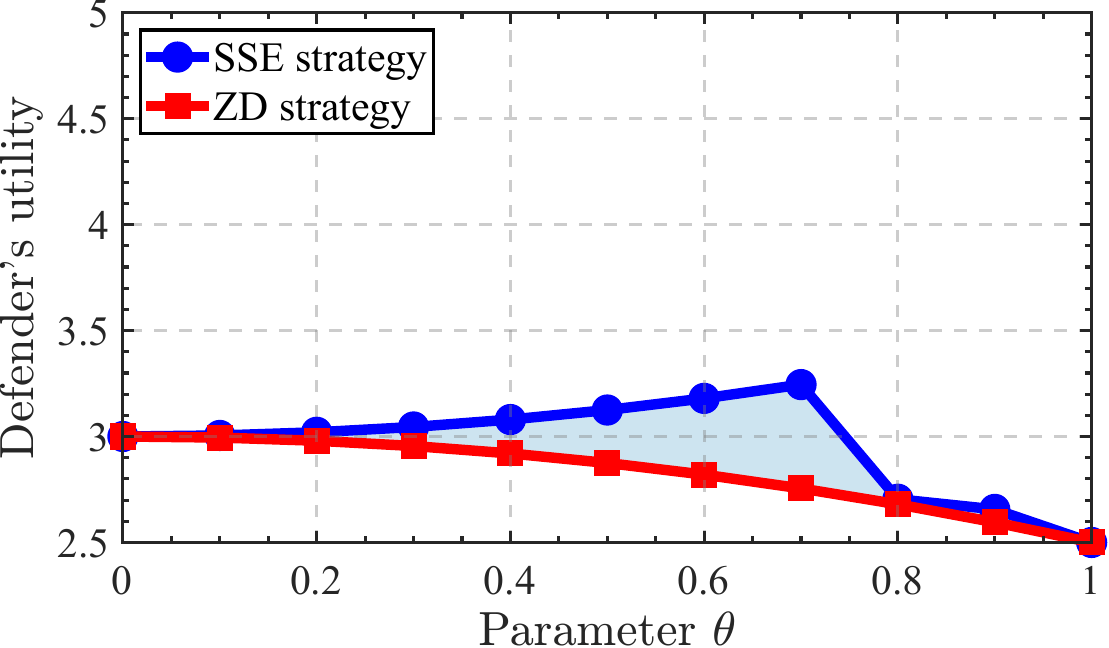}
\end{minipage}%
}%
\subfigure[$K=5,\zeta=2$]{\label{fi::1b}
\begin{minipage}[t]{0.32\linewidth}
\centering
\includegraphics[width=2in]{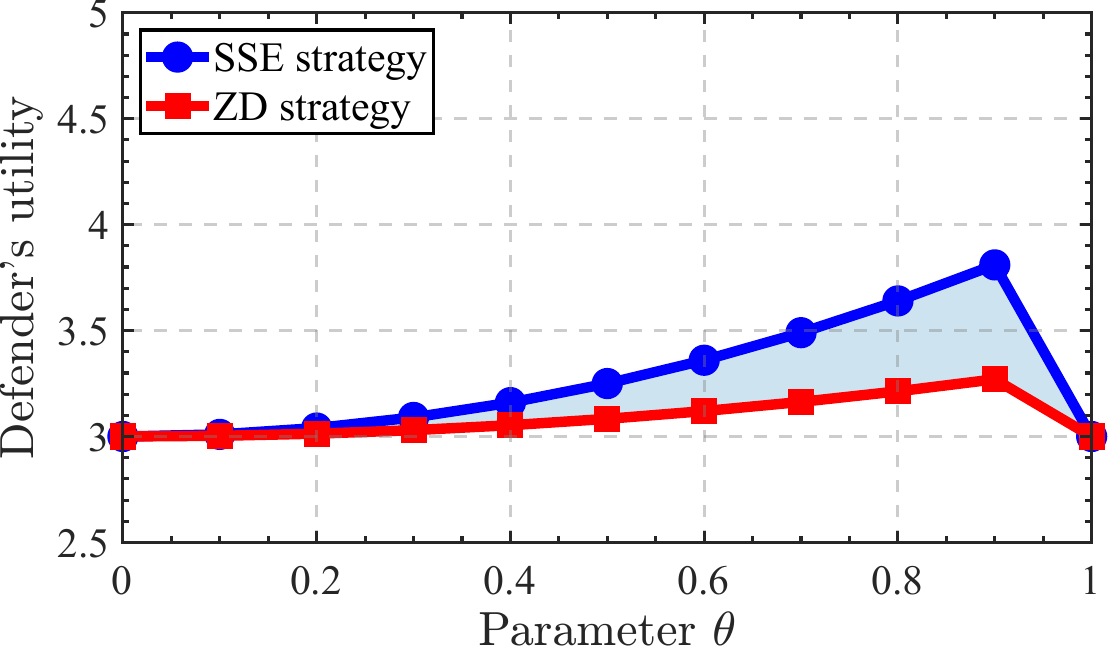}
\end{minipage}%
}%
\subfigure[$K=5,\zeta=3$]{\label{fi::1c}
\begin{minipage}[t]{0.32\linewidth}
\centering
\includegraphics[width=2in]{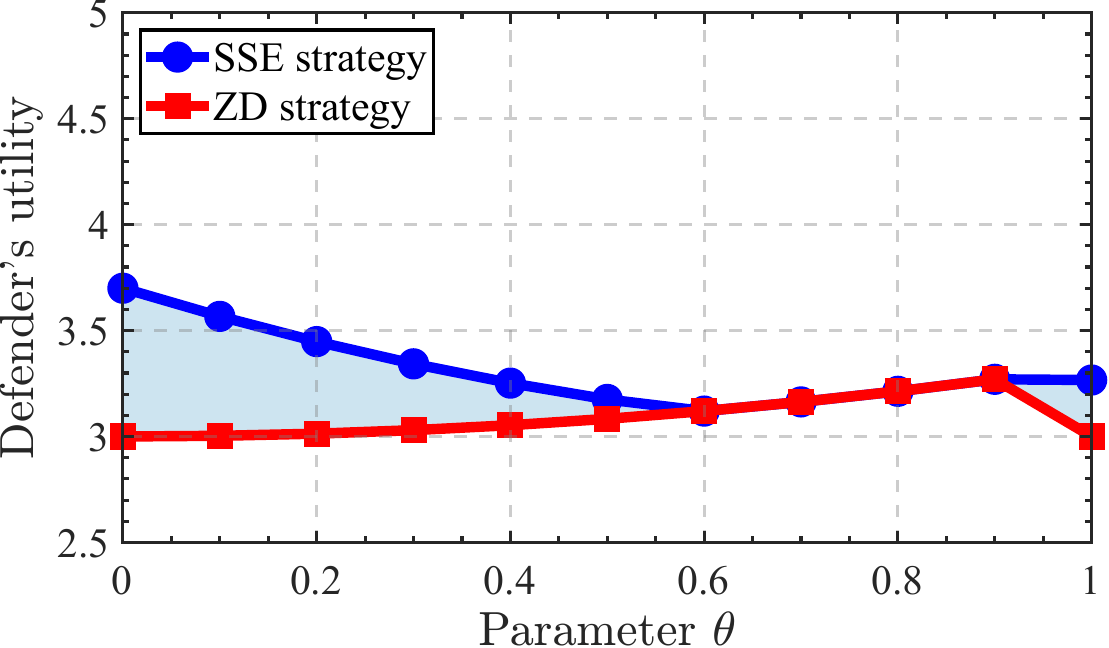}
\end{minipage}%
}%
\\
\vspace{-5pt}
\subfigure[$K=10,\zeta=1$]{\label{fi::1a}
\begin{minipage}[t]{0.32\linewidth}
\centering
\includegraphics[width=2in]{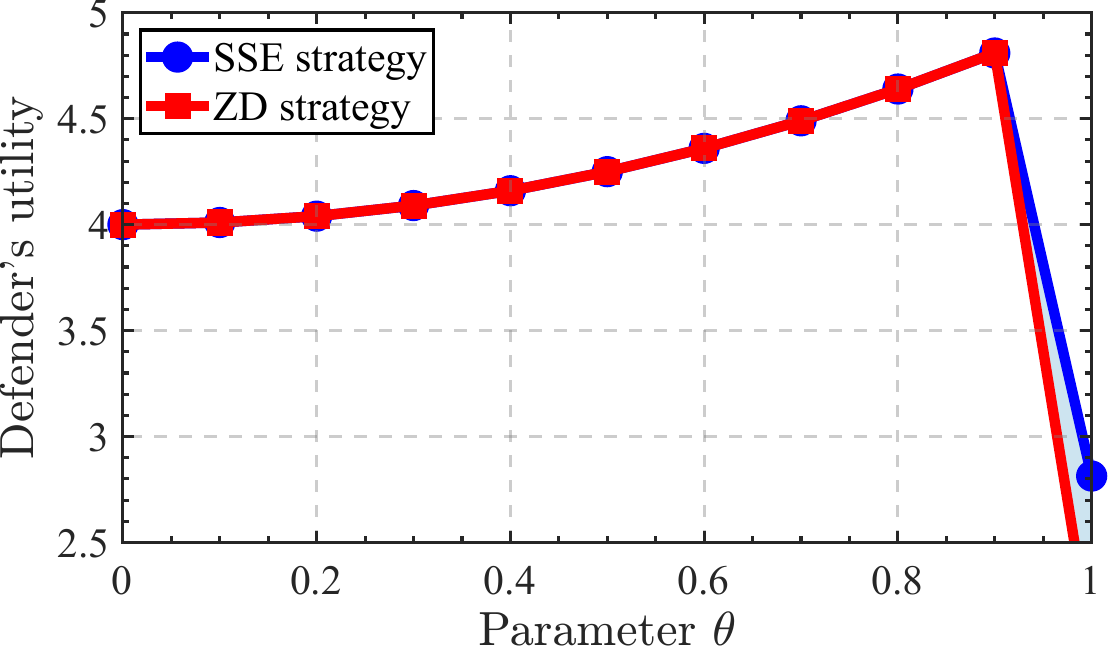}
\end{minipage}%
}%
\subfigure[$K=10,\zeta=2$]{\label{fi::1b}
\begin{minipage}[t]{0.32\linewidth}
\centering
\includegraphics[width=2in]{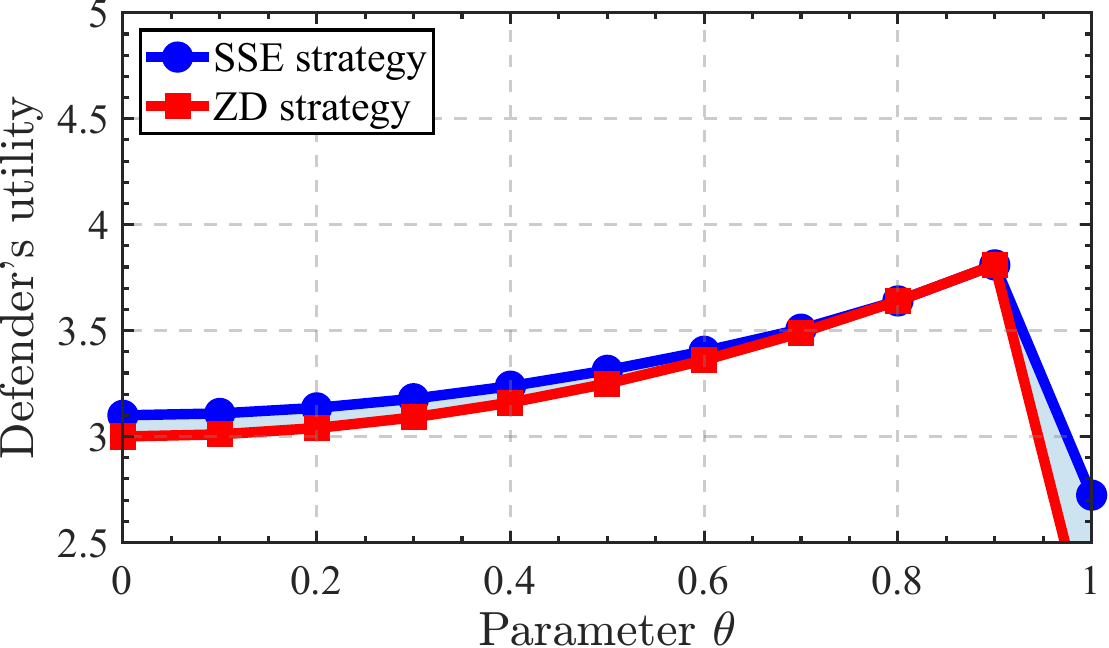}
\end{minipage}%
}%
\subfigure[$K=10,\zeta=3$]{\label{fi::1c}
\begin{minipage}[t]{0.32\linewidth}
\centering
\includegraphics[width=2in]{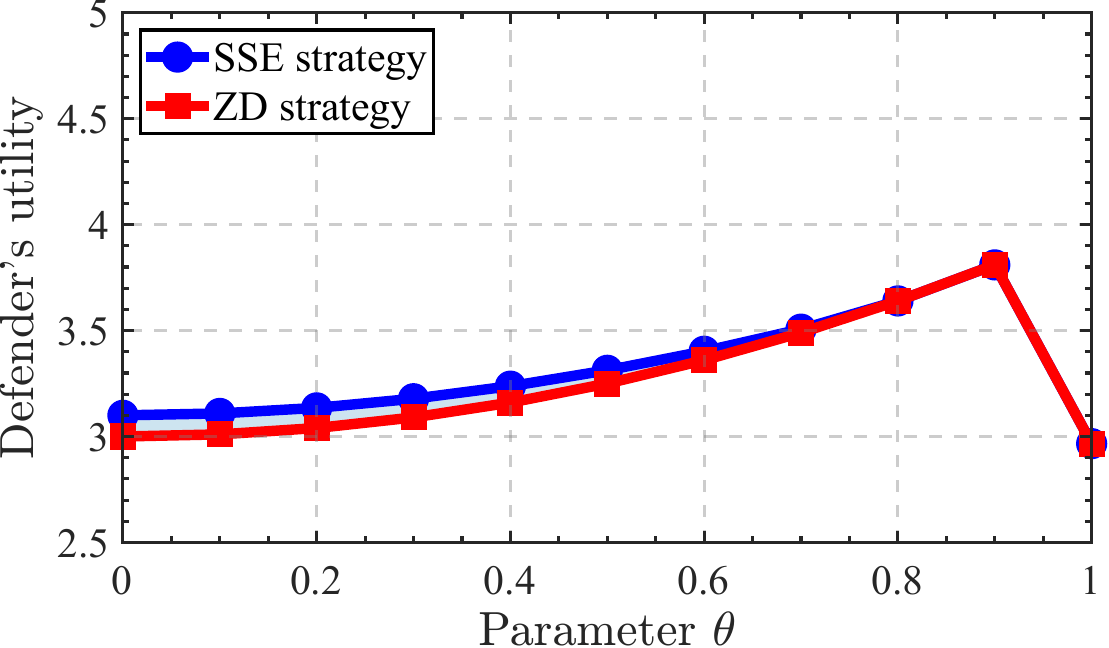}
\end{minipage}%
}%
\vspace{-5pt}
\caption{Performance comparison between the ZD strategy and the SSE strategy in an IoT system.
The red curve shows the defender’s utility achieved by the ZD strategy computed via Algorithm~\ref{Al::ZD construction}, while the blue curve corresponds to the defender’s utility under the SSE strategy obtained by solving \eqref{eq::optimization::compute::SSE}.
}
\label{fi::1}
\vspace{-10pt}
\end{figure*}

Thus, computing the exact SSE strategy via the mixed-integer program in~(\ref{eq::optimization::compute::SSE}) incurs a worst-case time complexity of $O(2^{K^3})$. The computational burden grows exponentially with respect to the number of targets $K$, making the exact computation of SSE strategies intractable for large-scale instances.


\textbf{Complexity of computing ZD:}
We analyze the computational complexity of the ZD strategy by examining the three steps outlined in Algorithm~\ref{Al::ZD construction}. Step 1 finds the linear parameters $\alpha$, $\beta$, and $\gamma$ of the ideal or optimal ZD strategies. 
To check for the existence of an ideal ZD strategy, one can solve the system of $K+4$ linear inequalities and equations in (\ref{eq::th::2}), which can be accomplished in polynomial time, e.g., with complexity $O(K^3)$ using standard linear programming techniques. If no ideal ZD strategy exists, the algorithm proceeds to solve the program (\ref{eq::optimization 2}) to obtain the optimal ZD strategies, with five variables: $\alpha$, $\beta$, $\gamma$, $u_d$, and $u_a$. 
This program involves $O(K)$ linear constraints arising from the union $\Lambda$, $O(K)$ constraints from the convex hull $\text{Conv}(G)$ (with at most $2K$ vertices), and the ZD equality $\alpha u_d + \beta u_a + \gamma = 0$.
Although the ZD equality introduces a non-convex coupling, the fixed number of variables ensures that the problem can be solved in time polynomial in the number of constraints.
Consequently, Step~1 can be solved with complexity $O(K^3)$.
Besides, Step~2 constructs the feasibility parameters $\{\phi_k\}_{k=1}^K$ using explicit formulas that involve $O(K)$ operations. 
Step~3 then computes the defender's ZD strategy $\pi_d$ iteratively, also with $O(K)$ operations. 
Since each step of Algorithm~\ref{Al::ZD construction} runs in polynomial time, the overall complexity of computing the ZD strategy is polynomial in $K$, more specifically, $O(K^3)$.

In contrast to the exponential complexity $O(2^{K^3})$ of computing the SSE strategy, the polynomial complexity $O(K^3)$ of computing the ZD strategy makes it a computationally efficient alternative. Thus, Algorithm \ref{Al::ZD construction} achieves comparable defense performance at a significantly lower computational cost, especially in large-scale security games with many targets.

\section{Experimental Validation}
\label{sec:experiment}

To verify our approach to construct ZD-driven MTD strategy, we validate our result in representative application scenarios including IoT systems and crowdsourcing systems. 


\subsection{IoT system}
\label{sec:iot-mtd-application}

We consider an IoT system \cite{azab2011chameleonsoft,feng2017signaling,wang2019moving} as shown in Fig. \ref{fi::Iot}, which is subject to persistent attacks. To protect critical resources over $K$ devices, the defender dynamically migrates the protection service across devices. At each stage, the defender selects a device $d_t$ to deploy the protection service, while the attacker chooses a device $a_t$ to attack. A successful defense means that the defender intercepts the attacker’s intrusion, i.e., $d_t = a_t$, while a successful attack occurs when the attacker evades the deployed protection, i.e., $d_t \neq a_t$
Specifically, $S>0$ denotes the defender’s gain from successful protection. The migration cost of the defender when relocating the protection service from device $i$ to device $j$ is denoted by $Y_{i,j} \geq 0$. 
For the attacker, $R>0$ represents the reward obtained from a successful attack, while $C_k>0$ denotes the corresponding attack cost on device $k$.
On this basis, the defender’s covered profit is $U_d^c(k) = S - \frac{1}{K}\sum_{i=1}^K Y_{i,k}$, where $\frac{1}{K}\sum_{i=1}^K Y_{i,k}$ represents the expected migration cost to device $k$, averaged over all possible prior locations  \cite{wang2019moving}. When the attacked device is not protected, the defender’s uncovered profit is $U_d^u(k) = -\frac{1}{K}\sum_{i=1}^K Y_{i,k}$, reflecting that the defender only incurs the amortized migration cost without obtaining any security gain.
Similarly, the attacker’s covered profit is $U_a^c(k) = - C_k$, which represents the attack cost. When the attacker successfully attacks an unprotected device, its uncovered profit is $U_a^u(k) = R -C_k$. To examine the utility differences between ZD strategies and SSE strategies, we introduce two parameters for different operational scenarios. The parameter $\theta \in [0,1]$ controls the defender’s migration cost profile $Y_{i,j}(\theta)$, capturing different levels of mobility and reconfiguration flexibility. The parameter $\zeta \in \{1,2,3\}$ indexes distinct attacker cost structures $C_k(\zeta)$, corresponding to varying attack capabilities and intensities.

As illustrated in Fig.~\ref{fi::1}, we compare the defender’s utility achieved by the ZD strategy via Algorithm~\ref{Al::ZD construction}, with that obtained under the SSE strategy. Although the SSE strategy consistently yields a higher utility, the deviation between the SSE and a ZD strategy remains small across all considered scenarios. This indicates that, despite the optimality of the SSE, the performance gap between the two strategies is limited. Notably, for certain parameter configurations, the ZD strategy achieves the same utility as the SSE strategy. These results show that adopting the proposed ZD strategy does not lead to a significant loss in defensive performance and, in some cases, can even preserve the optimal defense utility.

\begin{figure}[tbp]
\subfigure[Computation time in linear scale]{
 \includegraphics[width=3.3in]{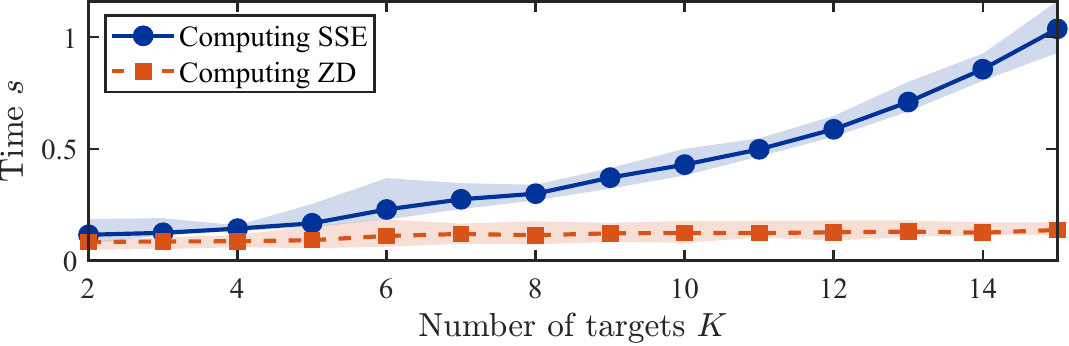}
 \vspace{-10pt}
 }\\
 \subfigure[Computation time in log scale]{
 \includegraphics[width=3.3in]{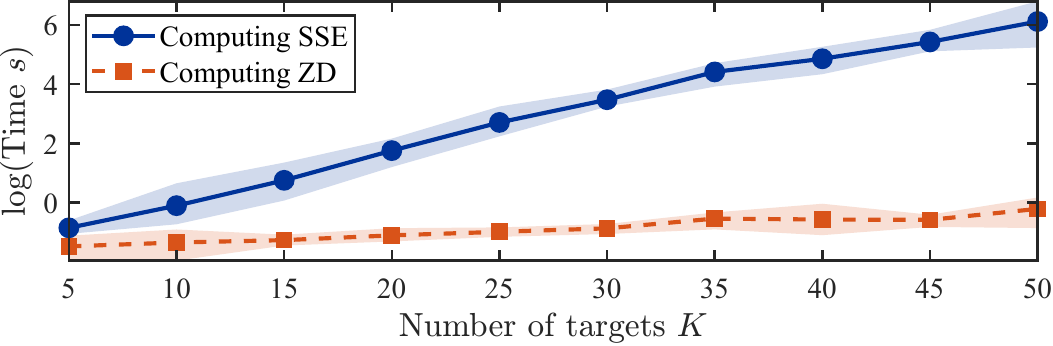}
 \vspace{-5pt}
 }
 \vspace{-5pt}
\caption{Average computation time of the ZD strategy and the SSE strategy.
The orange curve shows the average computation time to compute the ZD strategy via Algorithm~\ref{Al::ZD construction}, while the blue curve corresponds to the average computation time for solving the SSE strategy using \eqref{eq::optimization::compute::SSE}. The light orange and light blue shaded regions indicate the range of computation times (i.e., minimum to maximum values) over 50 independent runs, respectively.
}
\label{fi::2}
\vspace{-15pt}
\end{figure}

In Fig.~\ref{fi::2}, we further compare the computation time by the ZD strategy and that by the SSE strategy, where we randomly generate the game configurations and take the average computation time over $50$ independent cases. In Fig.~\ref{fi::2} (a),  it can be clearly observed that computing the SSE strategy requires substantially more time than computing the ZD strategy for $K \in [2,15]$. Moreover, as $K$ is large in Fig.~\ref{fi::2} (b), where the time on the vertical axis is taken as log, the computation time of the SSE strategy grows much more rapidly than that of the ZD strategy. Consequently, computing the SSE strategy becomes impractical for large-scale target problems, whereas the ZD strategy remains computationally efficient for maintaining defensive performance.

\subsection{Crowdsourcing system}

\begin{figure}[tbp]
\centering
 \includegraphics[width=3.5in]{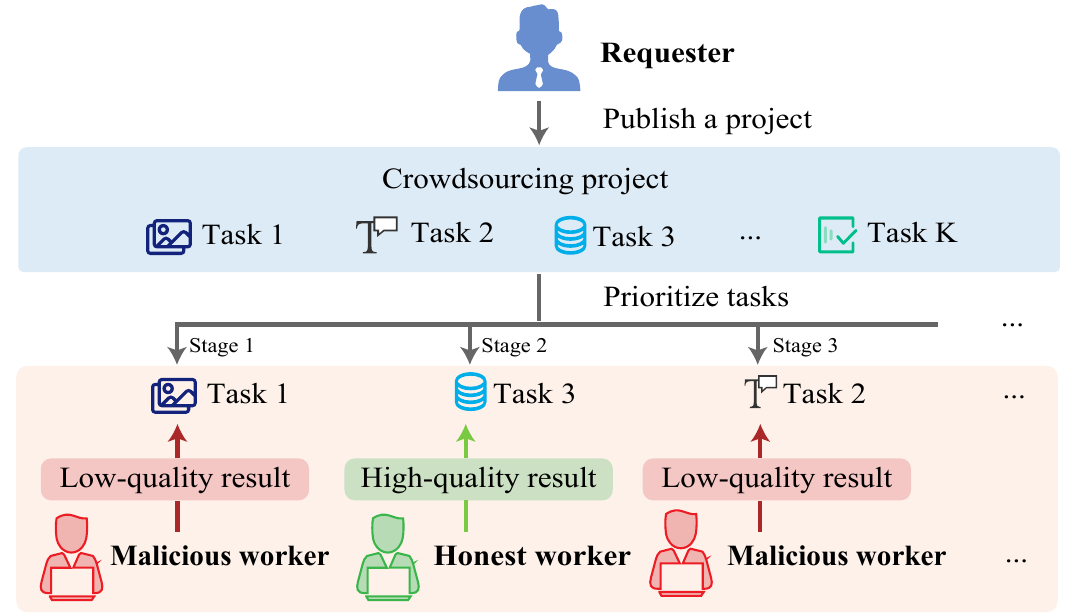}
 \vspace{-10pt}
\caption{Crowdsourcing system with honest and malicious workers}
\label{fi::crowdsourcing}
\vspace{-15pt}
\end{figure}

We consider a crowdsourcing system \cite{gadiraju2015understanding,hu2019solving,gao2021trustworker}, where a requester publishes a large-scale project consisting of $K$ tasks, such as image labeling, text translation, and data verification, as illustrated in Fig.~\ref{fi::crowdsourcing}. At each stage, the requester prioritize tasks and allocates verification resources to encourage the worker to focus on a task currently. The worker may be either honest (trustworthy) or malicious (untrustworthy). An honest worker faithfully follows the requester’s instructions and strives to deliver high-quality results. In contrast, a malicious worker seeks to hinder the progress of the project and obtain additional benefits by covertly submitting low-quality results or diverting effort to other tasks.
Specifically, let $R^r_k$ denote the requester’s reward when task $k$ is completed. If task $k$ is submitted with low quality due to malicious behavior, the requester gets an additional loss, $m_k$. Moreover, let $c$ denote the additional verification resource cost for the requester. For the worker, let $R^w_k$ and $\bar{R}^w_k$ denote rewards received by an honest worker and a malicious worker, respectively. Besides, let $a_k$ denote the extra benefit that a malicious worker can obtain by covertly diverting effort from task $k$ to other tasks.
Then, when the requester verifies task $k$, the requester’s covered profit is $U_d^c(k)= R_k^r-c$ if a high-quality result is submitted. Otherwise, $U_d^c(k)=R_k^r-m_k-c$.  When the requester does not verify task $k$, the requester’s uncovered profit reduces to $U_d^u(k) =-c$, reflecting that only the verification resource cost is incurred. For the worker, an honest worker receives a profit $U_a^c(k)=R^w_k$ when task $k$ is verified and accepted, and $U_a^u(k)=0$ otherwise. In contrast, a malicious worker obtains $U_a^c(k)=\bar{R}^w_k$, and $U_a^u(k) = \frac{1}{K}\sum_{i=1}^KR^w_i + a_k$.

\begin{figure*}[tbp]
\centering
\subfigure[Initially honest, period $50$]{\label{fi::31}
\begin{minipage}[t]{0.31\linewidth}
\centering
\includegraphics[width=2in]{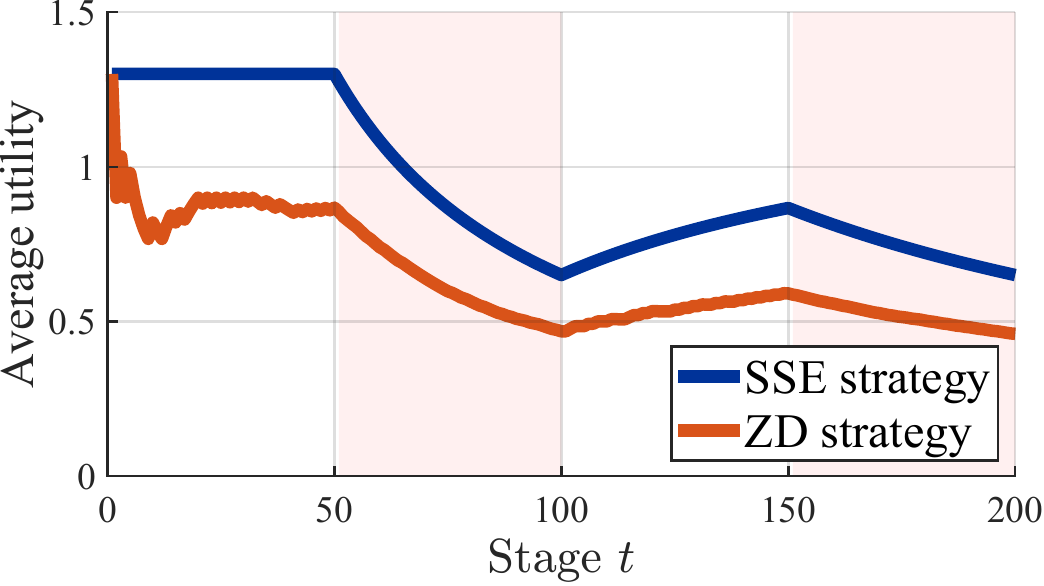}
\end{minipage}
}
\subfigure[Initially honest, period $20$]{\label{fi::32}
\begin{minipage}[t]{0.31\linewidth}
\centering
\includegraphics[width=2in]{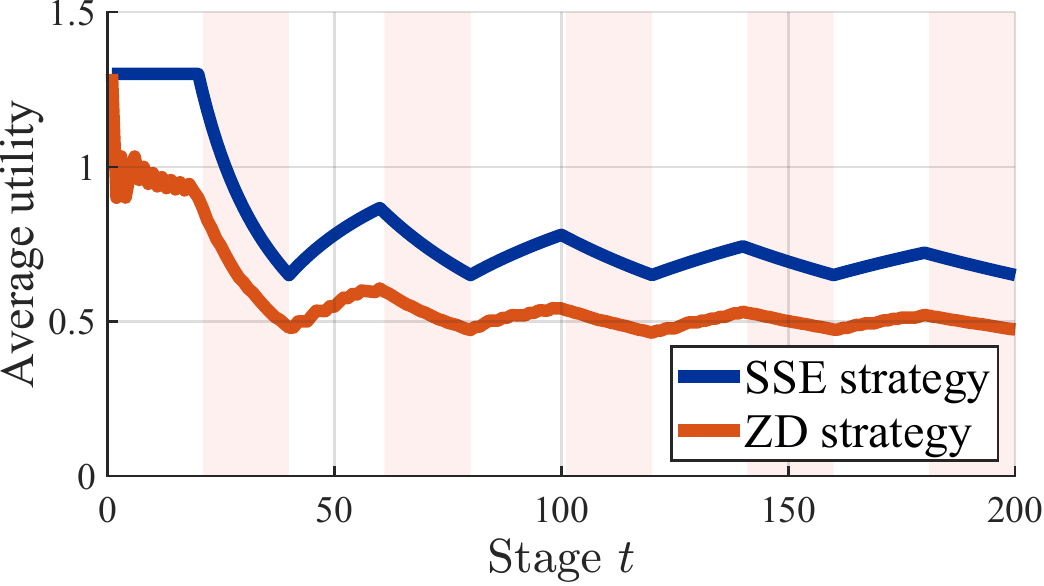}
\end{minipage}
}%
\centering
\subfigure[Initially honest, period $10$]{\label{fi::32}
\begin{minipage}[t]{0.31\linewidth}
\centering
\includegraphics[width=2in]{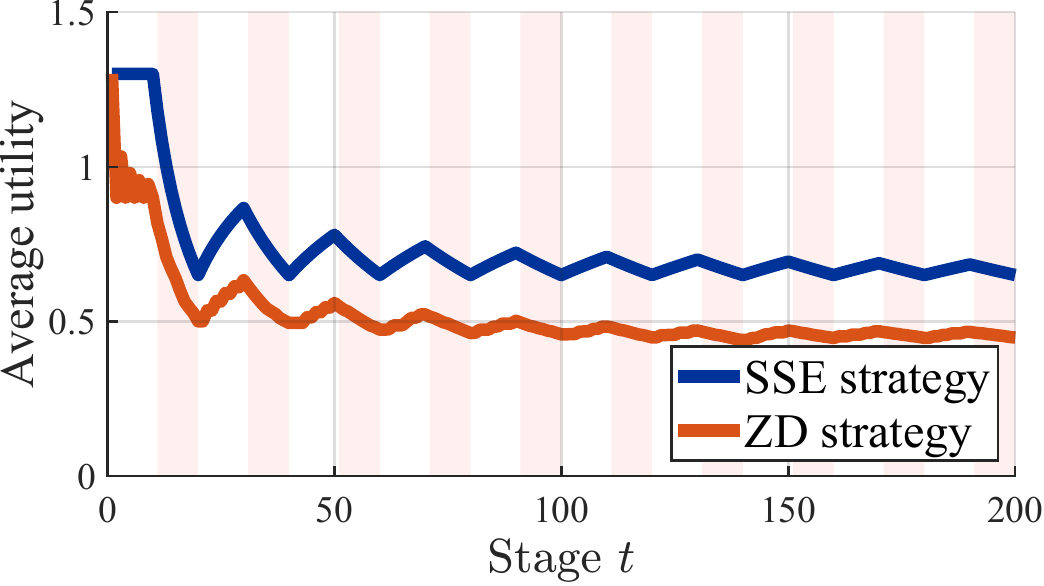}
\end{minipage}
}%
\\
\vspace{-7pt}
\centering
\subfigure[Initially malicious, period $50$]{\label{fi::33}
\begin{minipage}[t]{0.31\linewidth}
\centering
\includegraphics[width=2in]{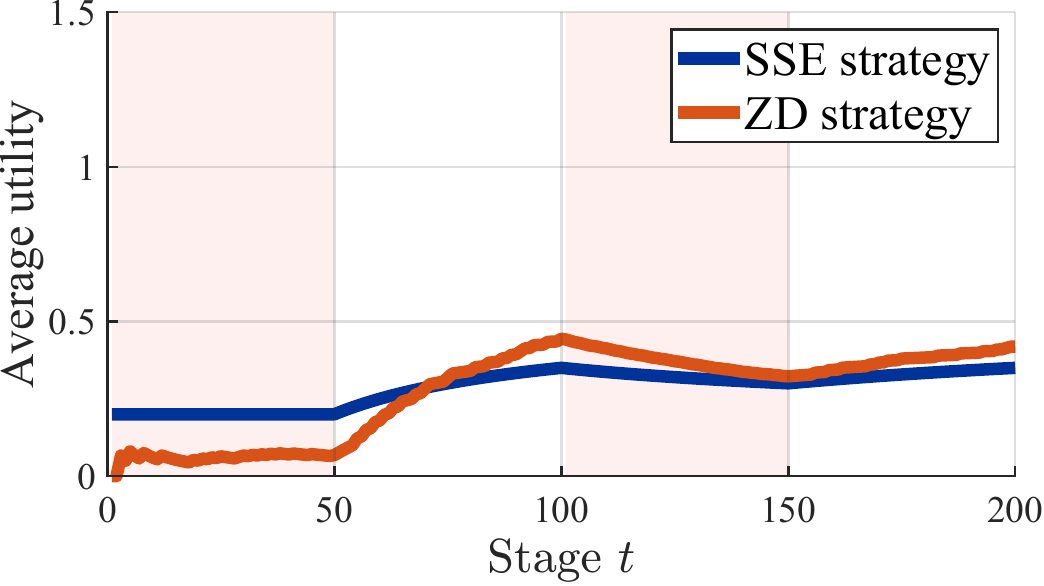}
\end{minipage}
\vspace{-7pt}
}
\centering
\subfigure[Initially malicious, period $20$]{\label{fi::34}
\begin{minipage}[t]{0.31\linewidth}
\centering
\includegraphics[width=2in]{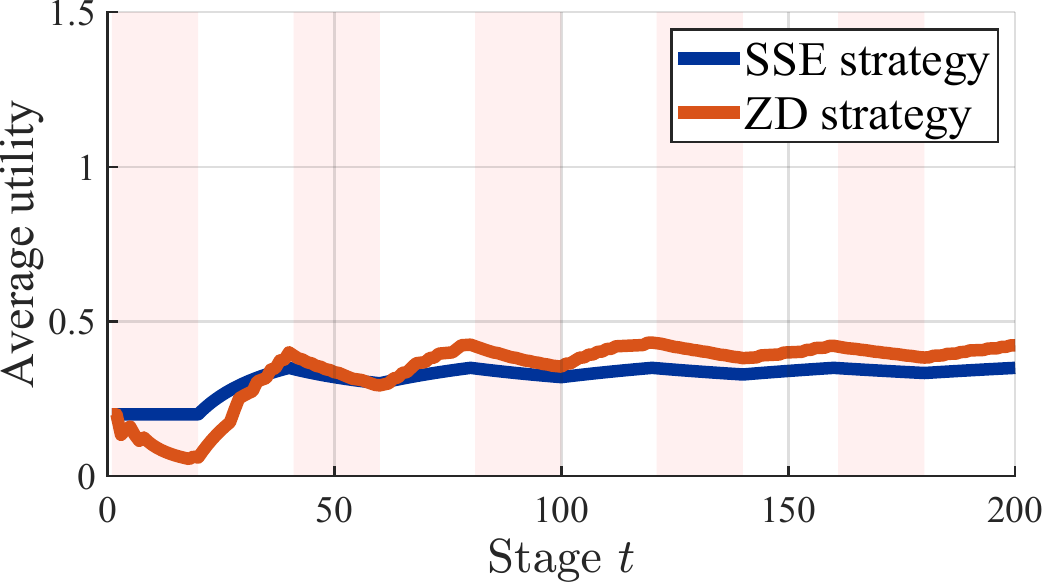}
\end{minipage}
\vspace{-7pt}
}%
\centering
\subfigure[Initially malicious, period $10$]{\label{fi::32}
\begin{minipage}[t]{0.31\linewidth}
\centering
\includegraphics[width=2in]{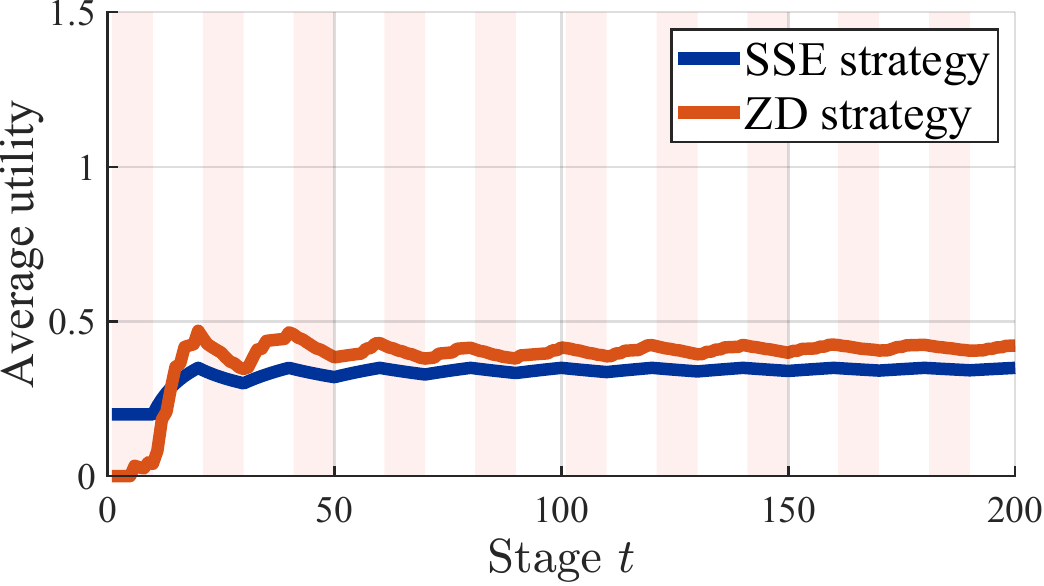}
\end{minipage}
}%
\vspace{-5pt}
\caption{Long-term average utility of the requester under the ZD and SSE strategies when interacting with a worker whose type switches periodically between honest and malicious. 
The orange curve represents the requester’s utility under the ZD strategy computed via Algorithm~\ref{Al::ZD construction}, 
while the blue curve corresponds to the utility under the SSE strategy obtained by solving \eqref{eq::optimization::compute::SSE}. 
Subfigures (a)–(c) consider scenarios where the worker is initially honest, and switches type with periods of $50$, $20$, and $10$, respectively. 
Subfigures (d)–(f) consider scenarios where the worker is initially malicious with the same switching periods. 
The light-red shaded regions indicate stages during which the worker is malicious, whereas the unshaded regions correspond to an honest worker.
}
\label{fi::3}
\vspace{-10pt}
\end{figure*}

As illustrated in Fig.~\ref{fi::3}, we present the requester’s long-term average utility under periodically switching worker types, where the worker alternates between honest and malicious behaviors over time.
As shown in Fig.~\ref{fi::3}(a)–(c), when the worker is initially honest, the ZD strategy results in smaller fluctuations in long-term average utility compared with the SSE strategy.
Moreover, as illustrated in Fig.~\ref{fi::3}(d)–(f), when the worker is initially malicious, the ZD strategy achieves higher long-term average utility than the SSE strategy.
This advantage arises from the unilateral enforcement property of the ZD strategy, allowing it to maintain stable and robust performance under dynamically changing adversarial behaviors.

\section{CONCLUSION}
This paper studied ZD strategies for constructing MTD strategies in repeated security games. We analyzed the existence and performance of ZD strategies and investigated their performance. To enable practical deployment, we developed programs for both the ideal ZD strategy and the optimal ZD strategy. Besides, we designed an algorithm to compute the proposed ZD strategies and proved their optimality guarantee. Compared with traditional SSE computation, the proposed approach significantly reduces computational complexity while maintaining comparable defensive performance. In the future, we plan to extend the proposed method to Markov games and repeated games with discounted long-term rewards, and scenarios with uncertainty and non-rational behaviors.

\bibliographystyle{IEEEtran} 
\bibliography{cite}

\appendix
\subsection{Proof of Lemma \ref{le::compute_SSE}}\label{app::le::1}
Suppose $\pi_a \in \mathbf{BR}(\pi_d)$. For any $\pi_d$, take $R_a(a|i,j)=\sum_{d}\pi_d(d|i,j)u_a(d,a)$ as the immediate profit by attacking target $a\in[K]$. 
According to \cite{puterman2014markov}, there exist $V_a^*$ and $Q^*:[K]\times[K]\to\mathbb{R}$ satisfying the Bellman optimality equation:
$V_a^* + Q^*(i,j) = \max_{a\in[K]}\left\{R_a(a|i,j) + \sum_{d}\pi_d(d|i,j)Q^*(d,a)\right\}$, for all $ i,j \in [K].$
Since $\pi_a$ is optimal and deterministic, for each $(i,j)$, let $a^*(i,j)$ satisfy $\pi_a(a^*(i,j)|i,j)=1$. Then
$
V_a^* + Q^*(i,j) = R_a(a^*(i,j)|i,j) + \sum_d \pi_d(d|i,j)Q^*(d,a^*(i,j)), $ and $
V_a^* + Q^*(i,j) \geq R_a(a|i,j) + \sum_d \pi_d(d|i,j)Q^*(d,a) $, for all  $a\neq a^*(i,j).$
Set $V_a=V_a^*$, $Q=Q^*$, and choose $Z$ large enough such that 
$Z > \left|R_a(a|i,j) + \sum_d \pi_d(d|i,j)Q(d,a) - V_a - Q(i,j)\right|.$
Similar to \cite{vorobeychik2012computing}, for $\pi_a(a|i,j)=1$, we have
$
V_a + Q(i,j) = \sum_d \pi_d(d|i,j)\left(u_a(d,a)+Q(d,a)\right).
$
For $\pi_a(k|i,j)=0$, the following inequality holds:
$$
\begin{aligned}
    V_a + Q(i,j) &\geq \sum\nolimits_d \pi_d(d|i,j)\left(u_a(d,k)+Q(d,k)\right), \\
V_a + Q(i,j) &\leq \sum\nolimits_d \pi_d(d|i,j)\left(u_a(d,k)+Q(d,k)\right) + Z.
\end{aligned}
$$
Similar to the above process, we can obtain 
$$
\begin{aligned}
V_d + W(i,j) &\leq \sum\nolimits_d \pi_d(d|i,j)\left(u_a(d,k)+W(d,k)\right) + Z.
\end{aligned}
$$
Since $\pi_a(k|i,j)\in\{0,1\}$, these conditions are equivalent to the condition in \eqref{eq::optimization::compute::SSE}. Noting that $V_d$ describes the defender's utility \cite{vorobeychik2012computing}, the defender's SSE strategy is a solution of  \eqref{eq::optimization::compute::SSE}. \hfill$\square$ 


\subsection{Proof of Theorem \ref{th::1}}\label{app::th::1}
   Since $\phi_K=0$, we have $\phi^{\min}_{-k}=0$, for $k<K$. Then (\ref{eq::ZD-parameter})  can be converted to 
\begin{subequations}
\begin{equation}\label{eq::ZD-parameter::1}
-\phi_k\leqslant\alpha  U_d^c(k)+\beta U_a^c(k)+\gamma\leqslant \phi^{\max}-\phi_k,
\end{equation}
\begin{equation}\label{eq::ZD-parameter::2}
0\leqslant\alpha  U_d^c(K)+\beta U_a^c(K)+\gamma\leqslant \phi^{\max},
\end{equation}
\begin{equation}\label{eq::ZD-parameter::3}
0\leqslant\alpha  U_d^u(k)+\beta U_a^u(k)+\gamma\leqslant\phi^{\max}-\phi^{\max}_{-k},
\end{equation}
\begin{equation}\label{eq::ZD-parameter::4}
-\phi^{\min}_{-K}\leqslant\alpha  U_d^u(K)+\beta U_a^u(K)+\gamma\leqslant \phi^{\max}-\phi^{\max}_{-K},
\end{equation}
\end{subequations}
where $k\neq K$. Inequality (\ref{eq::ZD-parameter::1})  is  equivalent to that there exists $\pi_d(i|k,k)\in[0,1]$, and $\sum\limits_{i}^{K-1}\pi_d(i|k,k)\leqslant 1$, such that
\begin{equation}\label{eq::pf::th1::1-1}
\alpha  U_d^c(k)+\beta U_a^c(k)+\gamma=\sum\limits_{i=1}^{K-1}\phi_i \pi_d(i|k,k)-\phi_k.
\end{equation}
Besides, inequality (\ref{eq::ZD-parameter::2}) is equivalent to the requirement that there exists $\pi_d(i|K,K)\in[0,1]$, $i=1,\dots,K$, and $\sum\limits_{i}^{K-1}\pi_d(i|K,K)\leqslant 1$, such that
\begin{equation}\label{eq::pf::th1::1-2}
\alpha  U_d^c(K)+\beta U_a^c(K)+\gamma=\sum\limits_{i=1}^{K-1}\phi_i \pi_d(i|K,K).
\end{equation}
Also, for $k=1,\dots,K-1$, inequality (\ref{eq::ZD-parameter::3}) is equivalent to
\begin{equation}\label{eq::pf::th1::5c1}
\begin{aligned}
-\phi_1&\leqslant\alpha  U_d^u(k)+\beta U_a^u(k)+\gamma\leqslant\phi^{\max}-\phi_1,\\
&\quad \quad\quad \quad\quad \quad\vdots\\
-\phi_{k-1}&\leqslant\alpha  U_d^u(k)+\beta U_a^u(k)+\gamma\leqslant\phi^{\max}-\phi_{k-1},\\
-\phi_{k+1}&\leqslant\alpha  U_d^u(k)+\beta U_a^u(k)+\gamma\leqslant\phi^{\max}-\phi_{k+1},\\
&\quad \quad\quad \quad\quad \quad\vdots\\
-\phi_{K-1}&\leqslant\alpha  U_d^u(k)+\beta U_a^u(k)+\gamma\leqslant\phi^{\max}-\phi_{K-1},
\end{aligned}
\end{equation}
and
\begin{equation}\label{eq::pf::th1::5c2}
\begin{aligned}
0&\leqslant\alpha  U_d^u(k)+\beta U_a^u(k)+\gamma\leqslant\phi^{\max}.
\end{aligned}
\end{equation}
Further, inequalities  (\ref{eq::pf::th1::5c1}) can be written as, for $j\neq k,K$,
$ -\phi_j\leqslant\alpha  U_d^u(k)+\beta U_a^u(k)+\gamma\leqslant\phi^{\max}-\phi_j.$ 
It is equivalent to that there exists $\pi_d(i|j,k)$, where $\sum\limits_{i}^{K-1}\pi_d(i|k,k)\leqslant 1$, such that
\begin{equation}\label{eq::pf::th1::1-3}
\alpha U_d^u(k)+\beta U_a^u(k)+\gamma= \sum\limits_{i=1}^{K-1}\phi_i \pi_d(i|j,k)-\phi_j.
\end{equation}
Besides, inequality (\ref{eq::pf::th1::5c2}) is equivalent to the condition that there exists $\pi_d(i|K,k)\in[0,1]$, $\sum\limits_{i}^{K-1}\pi_d(i|K,k)\leqslant 1$, such that
\begin{equation}\label{eq::pf::th1::1-4}
\begin{aligned}
\alpha U_d^u(k)+\beta U_a^u(k)+\gamma = \sum\limits_{i=1}^{K-1}\phi_i \pi_d(i|K,k).
\end{aligned}
\end{equation}
Moreover, inequality (\ref{eq::ZD-parameter::4}) is equivalent to 
$$
\begin{aligned}
-\phi_1&\leqslant\alpha  U_d^u(K)+\beta U_a^u(K)+\gamma\leqslant\phi^{\max}-\phi_1,\\
&\quad \quad\quad \quad\quad\quad\vdots\\
-\phi_{K-1}&\leqslant\alpha  U_d^u(K)+\beta U_a^u(K)+\gamma\leqslant\phi^{\max}-\phi_{K-1}.
\end{aligned}
$$
For $j\neq K$,  the above inequality is equivalent to that, there exists $\pi_d(i|j,K)\in[0,1]$, where $\sum\limits_{i}^{K\!-\!1}\pi_d(i|k,K)\leqslant 1$,  such that
\begin{equation}\label{eq::pf::th1::1-5}
\begin{aligned}
\alpha  U_d^u(K)+\beta U_a^u(K)+\gamma= \sum\limits_{i=1}^{K-1}\phi_i \pi_d(i|j,K)-\phi_j.
\end{aligned}
\end{equation}
Recalling (\ref{eq::pf::th1::1-1})(\ref{eq::pf::th1::1-2})(\ref{eq::pf::th1::1-3})-(\ref{eq::pf::th1::1-5}),  there exists $\pi_d$ such that 
$
\sum\limits_{k=1}^{K-1}\phi_k\left(\pi_d(k)-\hat{\pi}(k)\right)=\alpha \mathbf{S}^{d} +\beta \mathbf{S}^a +\gamma \mathbf{1}_{K^2},
$
where $\pi_d(K)=1-\sum\limits_{k=1}^{K-1}\pi_d(k)$.
By the equivalence property, (\ref{eq::th2222::ideal performance}) is sufficient and necessary for the existence of ZD strategies. \hfill$\square$

\subsection{Proof of Theorem \ref{le::ZDlemma2}}\label{app::th::2}
According to Definition~\ref{def::SSE}, $(\pi_d^{SSE}, \pi_a^{SSE})$ satisfies
$$
(\pi_d^{SSE}, \pi_a^{SSE})\in\mathop{\text{argmax}}\limits_{\pi_d, \pi_a\in {\textbf{BR}}({\pi_d})}\bar{u}_d(\pi_d,\pi_a).
$$
i.e., $\pi_d^{\mathrm{SSE}}$ is the strategy that attains the maximum expected utility for the defender, subject to the attacker playing a best response to the defender's strategy. Hence, for any feasible defender strategy $\pi_d$ and any best response strategy $\pi_a \in \mathbf{BR}(\pi_d)$ of the attacker, we must have
$$\bar{u}_d(\pi_d,\pi_a) \le \bar{u}_d(\pi_d^{\mathrm{SSE}},\pi_a^{\mathrm{SSE}}).$$
In particular, taking $\pi_d = \pi_d^{\mathrm{ZD}}$ and letting $\pi_a$ be any best response of the attacker to $\pi_d^{\mathrm{ZD}}$ (i.e., $\pi_a \in \mathbf{BR}(\pi_d^{\mathrm{ZD}})$), we immediately obtain
$\bar{u}_d(\pi_d^{ZD},\textbf{{BR}}(\pi_d^{ZD}))\leqslant \bar{u}_d(\pi_d^{SSE},\pi_a^{SSE}),$
which is exactly inequality~\eqref{eq::th2222::ideal performance}. \hfill$\square$

\subsection{Proof of Theorem \ref{th::2}}\label{app::th::3}

%

We aim to prove that if the condition in (\ref{eq::th::2}) holds, then there exists $\phi_1,\dots,\phi_{K-1},\phi_{K}\geqslant 0 $, where $\phi_{K}=0$, such that
\begin{equation}\label{eq::pf::th::2-total}
\begin{aligned}
-\phi_k&\leqslant\alpha U_d^c(k)+\beta U_a^c(k)+\gamma\leqslant \phi^{\max}-\phi_k,\\
-\phi^{\min}_{-k}&\leqslant\alpha U_d^u(k)+\beta U_a^u(k)+\gamma\leqslant\phi^{\max}-\phi^{\max}_{-k}.
\end{aligned}
\end{equation}
Take
\begin{equation}
\begin{aligned}
\phi_{K-1} & = \max\left\{|\alpha U_d^c (K)+\beta U_a^c (K)+\gamma|, \right.\\
&\quad\quad\quad\ \ \left.| \alpha U_d^u(K)+\beta U_a^u(K)+\gamma|\right\},\\
\phi_k&=|\alpha U_d^c (k)+\beta U_a^c (k)+\gamma|+\phi_{K-1}, k\neq 1,K, \\
\phi_1&= 2\sum\limits_{k=2}^{K-1}\phi_k+|\alpha U_d^u (1)+\beta U_a^u (1)+\gamma |.
\end{aligned}
\end{equation}
Thus, $\phi^{\max}=\phi_1$, $\phi^{\min}=\phi_K$, $\phi^{\min}_{-K}=\phi_{K-1}$, and $\phi^{\min}_{-k}=0$, for any $k\neq K$. Then,
$
-\phi_1\leqslant\alpha U_d^c(1)+\beta U_a^c(1)+\gamma\leqslant \phi^{\max}-\phi_1.
$
Besides, for $k=2,\dots,K$, notice that $\phi_1-\phi_{k}\geqslant \phi_{k}\geqslant |\alpha U_d^c(k)+\beta U_a^c(k)+\gamma|$. Then, we have
$
-\phi_k\leqslant\alpha U_d^c(k)+\beta U_a^c(k)+\gamma\leqslant \phi_1-\phi_k.
$
Also, for any $k$, $\phi_1-\phi_k\geqslant |\alpha U_d^u (1)+\beta U_a^u (1)+\gamma | $. Then 
$$
0\leqslant\alpha U_d^u(1)+\beta U_a^u(1)+\gamma\leqslant\phi^{\max}-\phi^{\max}_{-1}.
$$
For $k=2,\dots,K-1$, since $\alpha U_d^u (k)+\beta U_a^u (k)+\gamma = 0$, we have
$
0\leqslant\alpha U_d^u(k)+\beta U_a^u(k)+\gamma\leqslant\phi^{\max}-\phi^{\max}_{-k}.
$
Moreover, since $\phi^{\min}_{-K}= \phi_{K-1}\geqslant | \alpha U_d^u(K)+\beta U_a^u(K)+\gamma|$, we have
$$
-\phi^{\min}_{-k}\leqslant\alpha U_d^u(K)+\beta U_a^u(K)+\gamma\leqslant 0\leqslant \phi^{\max}-\phi^{\min}_{-K}.
$$
Therefore, inequalities in (\ref{eq::pf::th::2-total}) are satisfied.
According to Theorem \ref{th::1}, there exists a ZD strategy 
$\pi_d^{ZD}$ which enforces the linear relation, $\alpha \bar{u}_d(\pi_d^{ZD},\pi_a)+\beta \bar{u}_a(\pi_d^{ZD},\pi_a)+\gamma=0$.


Without loss of generality, take $U_d^c (1)=\max\{U_d^c (1),\dots,U_d^c (K)\}$. Then $u_d(d,a) = x_aU_d^c(a)+(1-x_a)U_d^u \leqslant U_d^c(a)\leqslant U_d^c(1)$. If the attacker always chooses target $1$, i.e. $\pi_a(1|j,k)=1$, for any $j,k$, then
$\mathbb{E}\left[u_d(d_t,a_t)\right]=\mathbb{E}\left[x_aU_d^c(a)+(1-x_a)U_d^u(a)\right]=\mathbb{E}\left[x_1U_d^c(1)+(1-x_1)U_d^u(1)\right]=\mathbb{E}\left[1-x_1\right](U_d^u(1)-U_d^c(1))+U_d^c(1)=(1-\mathbb{E}\left[x_1\right])(U_d^u(1)-U_d^c(1))+U_d^c(1)$. Noticing that $\mathbb{E}\left[x_1\right]=1$, then
$\lim_{T\to\infty}\mathbb{E}\left[u_d(d_t,a_t)\right]=U_d^c(1)$, and $\bar{u}_d(\pi_d^{ZD},\pi_a) = U_d^c(1)$. Thus, $U_d^c(1)$ is the maximum value for the defender and can be achieved when the attacker always attacks target $1$. 

Suppose $\beta\neq0$. Due to the linear relation,  $ \bar{u}_a(\pi_d^{ZD},\pi_a)=-\frac{\alpha}{\beta} \bar{u}_d(\pi_d^{ZD},\pi_a)-\frac{\gamma}{\beta}$ and $-\frac{\alpha}{\beta}\geq 0$, which means that the attacker's utility has an non-negative correlation with the defender's utility. Given the ZD strategy $\pi_d^{ZD}$, if the attacker aims to maximize its own utility, the attacker also maximizes $ \bar{u}_d(\pi_d^{ZD},\pi_a)$. The results hold for $\beta = 0$. Thus, the strategy that attacker always attacks target $1$ is a BR strategy of the attacker, and $\bar{u}_d(\pi_d^{ZD},\textbf{{BR}}(\pi_d^{ZD}))= U_d^c(1)$. According to Theorem \ref{le::ZDlemma2}, since  $U_d(\pi_d^{SSE},\pi_a^{SSE})\leqslant U_d^c (1)$, $U_d(\pi_d^{ZD},\pi_a^{{BR}}(\pi_d))= U_d(\pi_d^{SSE},\pi_a^{SSE})$. \hfill$\square$

\subsection{Proof of Theorem \ref{th::optimal_ZD}}\label{app::th::4}
Let $(\alpha^*,\beta^*,\gamma^*,u_d^*,u_a^*)$ be an optimal solution of the program \eqref{eq::optimization 2}. We aim to show $\alpha^*,\beta^*,\gamma^*$ are the linear parameters corresponding to the optimal ZD strategy. 
Since $(\alpha^*,\beta^*,\gamma^*) \in \Lambda$, there exist $i_1,i_2 \in [K]$ such that $(\alpha^*,\beta^*,\gamma^*) \in \Lambda(i_1,i_2)$. According to Theorem \ref{th::1}, there are nonnegative parameters $\phi_1,\dots,\phi_{K-1}$ with $\phi_K=0$ such that inequalities \eqref{eq::ZD-parameter} hold. Specifically, for $k \notin \{i_1,i_2\}$, set $\phi_k=0$; then the equalities in $\Lambda(i_1,i_2)$ imply that the inequalities in \eqref{eq::ZD-parameter} are satisfied. For $i_1$ and $i_2$, choose a sufficiently large $\phi^{\max}>0$ and define
$        \phi_{i_1} = \phi^{\max} - (\alpha^* U_d^c(i_1) + \beta^* U_a^c(i_1) + \gamma^*), $ $
        \phi_{i_2} = \phi^{\max} - (\alpha^* U_d^c(i_2) + \beta^* U_a^c(i_2) + \gamma^*).$
    The conditions in $\Lambda(i_1,i_2)$ guarantee that $\phi_{i_1},\phi_{i_2} \geq 0$, and a direct check shows that inequalities \eqref{eq::ZD-parameter} for $k=i_1,i_2$ are satisfied. Hence, by Theorem \ref{th::1}, there exists a ZD strategy $\pi_d^* $ such that for any $\pi_a$,
    \begin{equation}\label{eq:linear-relation-optimal}
        \alpha^* \bar{u}_d(\pi_d^*,\pi_a) + \beta^* \bar{u}_a(\pi_d^*,\pi_a) + \gamma^* = 0.
    \end{equation}

    Besides, the constraints of \eqref{eq::optimization 2} require $(u_d^*,u_a^*) \in \operatorname{Conv}(G)$ and $\alpha^* u_d^* + \beta^* u_a^* + \gamma^* = 0$.   
    An attacker strategy $\pi_a^{\mathrm{BR}}$ can be constructed to be a best response to $\pi_d^*$ and yields exactly the utility pair $(u_d^*,u_a^*)$ according to \cite{press2012iterated}. Thus, 
    \[
        \bar{u}_d(\pi_d^*,\pi_a^{\mathrm{BR}}) = u_d^*, \quad
        \bar{u}_a(\pi_d^*,\pi_a^{\mathrm{BR}}) = u_a^*.
    \]
    Since $(\alpha^*,\beta^*,\gamma^*,u_d^*,u_a^*)$ is an optimal solution of \eqref{eq::optimization 2} and every ZD strategy together with a best response yields a feasible point of \eqref{eq::optimization 2}, the value $u_d^*$ is the maximum defender utility achievable by any ZD strategy when the attacker plays a best response. Hence $\pi_d^*$ is an optimal ZD strategy, and its linear parameters are given by $(\alpha^*,\beta^*,\gamma^*)$. \hfill$\square$

\subsection{Proof of Theorem \ref{pro::algorithmZD}}\label{app::th::5}

Due to the construction of $\{\phi_k\}_{k=1}^K$, we have $\phi_1\geqslant\phi_2,\dots,\phi_{K-2}\geqslant\phi_{K-1} \geqslant\phi_K=0$. Thus, $\phi^{\max}=\phi_1$, $\phi^{\min}=\phi_{K}$, $\phi^{\max}_{-k}=\phi_{1}$, for $k\neq 1$, $\phi_{\max,1}=\phi_2$, $\phi^{\min}_{-k}=\phi_K$, for $k\neq K$, and $\phi^{\min}_{-k}=\phi_{K-1}$. For any $k\neq 1,K$, since $\phi_k=|\alpha U_d^c (k)+\beta U_a^c (k)+\gamma|+\phi_{K-1}$, we have 
$-\phi_k\leqslant\alpha U_d^c (k)+\beta U_a^c (k)+\gamma.$
Besides, $\phi^{\max}-\phi_{k}=\phi_1-\phi_{k}=2\sum\limits_{i=2}^{K-1}\phi_i-\phi_k+|\alpha U_d^u (1)+\beta U_a^u (1)+\gamma |=\phi_k+2\sum\limits_{i=2,i\neq k}^{K-1}\phi_i+|\alpha U_d^u (1)+\beta U_a^u (1)+\gamma |\geqslant \phi_k\geqslant\alpha U_d^c (k)+\beta U_a^c (k)+\gamma$.
Moreover, since $\phi_1= 2\sum\limits_{k=2}^{K-1}\phi_k+|\alpha U_d^u (1)+\beta U_a^u (1)+\gamma |+|\alpha U_d^c (1)+\beta U_a^c (1)+\gamma |$, we have $-\phi_{1}\leqslant \alpha U_d^c (1)+\beta U_a^c (1)+\gamma\leqslant 0$. 
Also, since $\phi_{K-1}  = \max\left\{|\alpha U_d^c (K)+\beta U_a^c (K)+\gamma|,  | \alpha U_d^u(K)+\beta U_a^u(K)+\gamma|\right\},$ we have $-\phi_{K-1}\leqslant \alpha U_d^c (1)+\beta U_a^c (1)+\gamma\leqslant \phi^{\max}-\phi_1$. Thus, for any $k$, 
$$-\phi_{k}\leqslant \alpha U_d^c (k)+\beta U_a^c (k)+\gamma\leqslant \phi^{\max}-\phi_k.$$

Besides, for $k\neq 1,K$, we have $\phi_{k}=0$ and $\phi^{\max}-\phi^{\max}_{-k}=0$. Also, for $k=1$, $\phi^{\min}_{-1}=\phi_{K}=0$ and $\phi^{\max}-\phi_{\max.1}=\phi_1-\phi_{\max,1}=2\sum\limits_{k=2}^{K-1}\phi_k-\phi_{\max,2}+|\alpha U_d^u (1)+\beta U_a^u (1)+\gamma |+|\alpha U_d^c (1)+\beta U_a^c (1)+\gamma |\geqslant \alpha U_d^u (1)+\beta U_a^u (1)+\gamma$. Also, for $k=K$, $-\phi^{\min}_{-k}=-\phi_{K-1}\leqslant U_d^u(K)+\beta U_a^u(K)+\gamma$ and $\phi^{\max}-\phi^{\max}_{-k}$. Therefore, for any $k$, 
$$-\phi^{\min}_{-k}\leqslant\alpha  U_d^u(k)+\beta U_a^u(k)+\gamma\leqslant\phi^{\max}-\phi^{\max}_{-k}.$$
Thus, $\{\phi_k\}_{k=1}^K$ is a solution of (\ref{eq::ZD-parameter}).

Moreover, due to the construction of $$\pi_d (k)\!\!=\!\!\left[\!\!\frac{\alpha \mathbf{S}^{D}\! \!+\!\!\beta \mathbf{S}^A \!\!+\!\!\gamma \mathbf{1}_{K^2}\!\!-\!\!\!\sum\limits_{i=k+1}^{K-1}\!\phi_i\left(\pi_d(i)\!\!-\!\hat{\pi}(i)\right)\!\!-\!\!\omega_k}{\phi_k}\!\!+\!\!\hat{\pi}(k)\!\!\right]^+\!\!\!,$$ we have $\pi_d(k|i,j)\geqslant0$. 
For $k=K-1$, $$\pi_d(K-1)=\!\left[\!\frac{\alpha \mathbf{S}^{d} \!+\!\beta \mathbf{S}^a \!+\!\gamma \mathbf{1}_{K^2}-\omega_1}{\phi_{K-1}}\!+\!\hat{\pi}(K-1)\right]^+.$$ Since $\phi_{K-1}  = \max\{|\alpha U_d^c (K)+\beta U_a^c (K)+\gamma|, $ $ | \alpha U_d^u(K)+\beta U_a^u(K)+\gamma|\}$, for $i\neq K-1, j\neq i$, there exists $\omega_{K-1}$ such that $\pi_d(K-1|i,i)\leqslant \left|\frac{\alpha U_d^c(i) \!+\!\beta U_a^c(i) \!+\!\gamma -\omega_{K-1}}{\phi_{K-1}}\right|\leqslant1$ and $\pi_d(K-1|i,j)\leqslant \left|\frac{\alpha U_d^u(i) \!+\!\beta U_a^u(i) \!+\!\gamma -\omega_{K-1}}{\phi_{K-1}}\right|\leqslant1$. 
Also, $\pi_d(K-1|,K-1,K-1)\leqslant \left|\frac{\alpha U_d^c(K-1) \!+\!\beta U_a^c(K-1) \!+\!\gamma -\omega_{K-1}}{\phi_{K-1}}+1\right|$, and $\pi_d(K-1|,K-1,j)\leqslant \left|\frac{\alpha U_d^u(K-1) \!+\!\beta U_a^u(K-1) \!+\!\gamma -\omega_{K-1}}{\phi_{K-1}}+1\right|$. Therefore, $\pi_d(K-1|i,j)\leqslant1$ for $i,j\in[K]$. Moreover, consider $\pi_d(k+1|i,j)\leqslant1$, $k\leqslant K-1$, we need to show $\pi_d(k|i,j)\leqslant1$ in the following.
$$\begin{aligned}\pi_d(k)\leqslant&\frac{\alpha \mathbf{S}^{D}\! \!\!+\!\!\beta \mathbf{S}^A \!\!+\!\!\gamma \mathbf{1}_{K^2}\!\!-\!\!\!\!\sum\limits_{i=k+1}^{K-1}\!\!\phi_i\left(\pi_d(i)\!\!-\!\!\hat{\pi}(i)\!\right)\!-\!\omega_k}{\phi_k}\!\!+\!\!\hat{\pi}(k)\\
=&\frac{\phi_{k+1}}{\phi_{k}}\!\frac{\alpha \mathbf{S}^{D} \!\!\!+\!\!\!\beta \mathbf{S}^A \!\!\!+\!\!\!\gamma \mathbf{1}_{K^2}\!\!\!-\!\!\!\sum\limits_{i=k+2}^{K-1}\!\!\phi_i\left(\pi_d(i)\!\!-\!\!\hat{\pi}(i)\right)\!-\!\omega_{k+1}}{\phi_{k+1}}\\
&-\!\frac{\phi_{k\!+\!1}}{\phi_k}\hat{\pi}(k\!+\!1)\!\!-\!\!\frac{\phi_{k\!+\!1}}{\phi_k}(\pi_d(k\!+\!1)\!-\!\hat{\pi}(k\!+\!1))\\
&+\!\hat{\pi}(k)\!+\!\frac{\omega_{k\!+\!1}\!-\!\omega_k}{\phi_k}\\
\leqslant&\frac{\phi_{k+1}}{\phi_{k}}\pi_d(k+1)+\hat{\pi}(k)+\frac{\omega_{k+1}-\omega_k}{\phi_k}\\
&-\frac{\phi_{k+1}}{\phi_k}\hat{\pi}(k+1)-\frac{\phi_{k+1}}{\phi_k}(\pi_d(k+1)-\hat{\pi}(k+1))\\
=&\hat{\pi}(k)+\frac{\omega_{k+1}-\omega_k}{\phi_k}.\end{aligned}$$ Thus, there exists $\omega_k$ such that $\pi_d(k)\leqslant1$. 
As a result, $\{\phi_k\}_{k=1}^K$ and $\pi_d$ are the solution of (\ref{eq::ZD-def-muliti}). \hfill$\square$

\end{document}